\pdfoutput=1

\documentclass[epj]{svjour}

\usepackage{graphicx}
\usepackage{dcolumn}
\usepackage{bm}

\usepackage{subfig}

\usepackage{array} 
\usepackage{lineno}
\usepackage{amsmath} 
\usepackage{amssymb}
\usepackage[numbers]{natbib}
\usepackage{comment}
\usepackage{stackrel}
\usepackage{mathtools}
\usepackage{multirow}
\usepackage{tabularx}
\usepackage{array}
\usepackage{hyperref}
\usepackage[thinlines]{easytable}
\usepackage{pifont}
\newcommand{\cmark}{\text{\ding{51}}}
\newcommand{\xmark}{\text{\ding{55}}}

\usepackage{makecell}

\usepackage{tikz}
\usetikzlibrary{arrows,positioning} 
\usepackage{varwidth}

\tikzset{
	>=stealth',
	box/.style={
		rectangle,
		rounded corners,
		dashed,
		draw=black, very thick,
		minimum height=2em,
		text centered,
		execute at begin node={\begin{varwidth}{28em}},
		execute at end node={\end{varwidth}}},
	solidbox/.style={
		rectangle,
		rounded corners,
		draw=black, very thick,
		minimum height=2em,
		text centered,
		execute at begin node={\begin{varwidth}{28em}},
			execute at end node={\end{varwidth}}},
	fw_arrow/.style={
		->,
		thick,
		shorten <=2pt,
		shorten >=2pt,},
	bw_arrow/.style={
		<-,
		thick,
		shorten <=2pt,
		shorten >=2pt,}
}

\allowdisplaybreaks

\begin{document}

\title{A unified perspective on modified Poisson likelihoods for limited Monte Carlo data}
\author{Thorsten Gl\"usenkamp
	\thanks{thorsten.gluesenkamp@fau.de}%
}                     
\institute{Erlangen Centre for Astroparticle Physics (ECAP), Erlangen}

\date{Received: date / Revised version: date}
%
\abstract
{
	Counting experiments often rely on Monte Carlo simulations for predictions of Poisson expectations. The accompanying uncertainty from the finite Monte Carlo sample size can be incorporated into parameter estimation by modifying the Poisson likelihood. We first review previous Frequentist methods of this type by Barlow et al, Bohm et al, and Chirkin, as well as recently proposed probabilistic methods by the author and Arg\"uelles et al. We show that all these approaches can be understood in a unified way: they all approximate the underlying probability distribution of the sum of weights in a given bin, the compound Poisson distribution (CPD). The Probabilistic methods marginalize the Poisson mean with a distribution that approximates the CPD, while the Frequentist counterparts optimize the same integrand treating the mean as a nuisance parameter. With this viewpoint we can motivate three new probabilistic likelihoods based on generalized gamma-Poisson mixture distributions which we derive in analytic form.  Afterwards, we test old and new formulas in different parameter estimation settings consisting of a "background" and "signal" dataset. The probablistic counterpart to the Ansatz by Barlow et al. outperforms all other existing approaches in various scenarios. We further find a surprising outcome: usage of the exact CPD is actually bad for parameter estimation. A continuous approximation performs much better and in principle allows to perform bias-free inference at any level of simulated livetime if the first two moments of the CPD of each dataset are known exactly. Finally, we also discuss the situation where new Monte Carlo simulation is produced for a given parameter choice which leads to fluctuations in the computed likelihood value. Two of the new formulas allow to include this Poisson uncertainty directly into the likelihood which substantially decreases these fluctuations.
} 
\maketitle

\tableofcontents 

\section{Introduction}
\label{sec:introduction}

\begin{figure}
\centering
\subfloat[Equal weights]{{\includegraphics{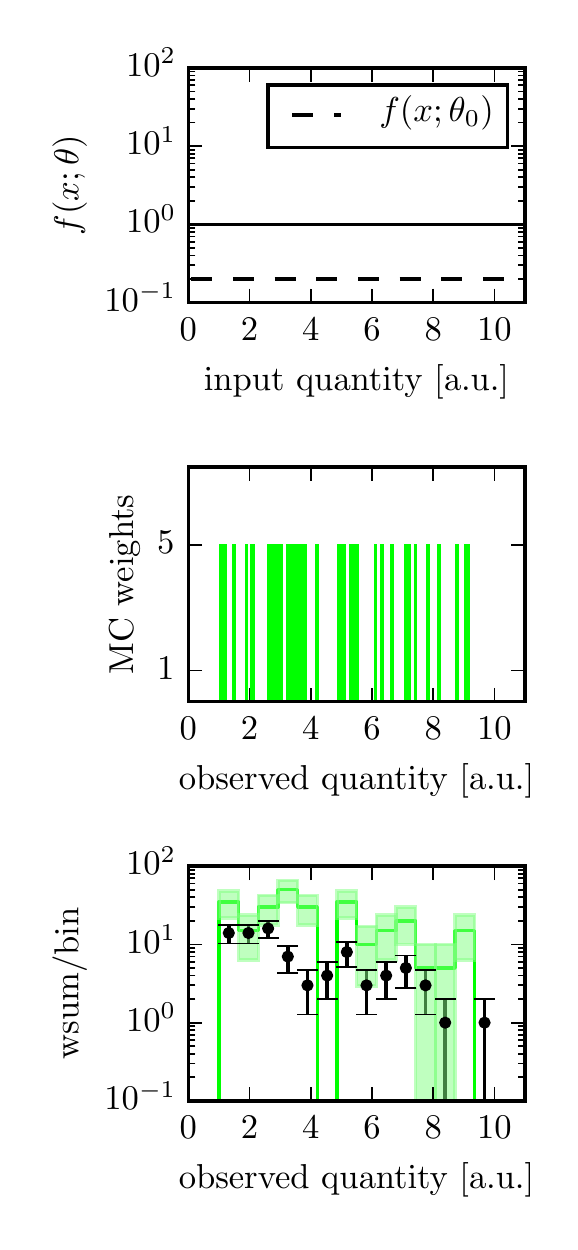}}}
\subfloat[Re-weighted events]{{\includegraphics{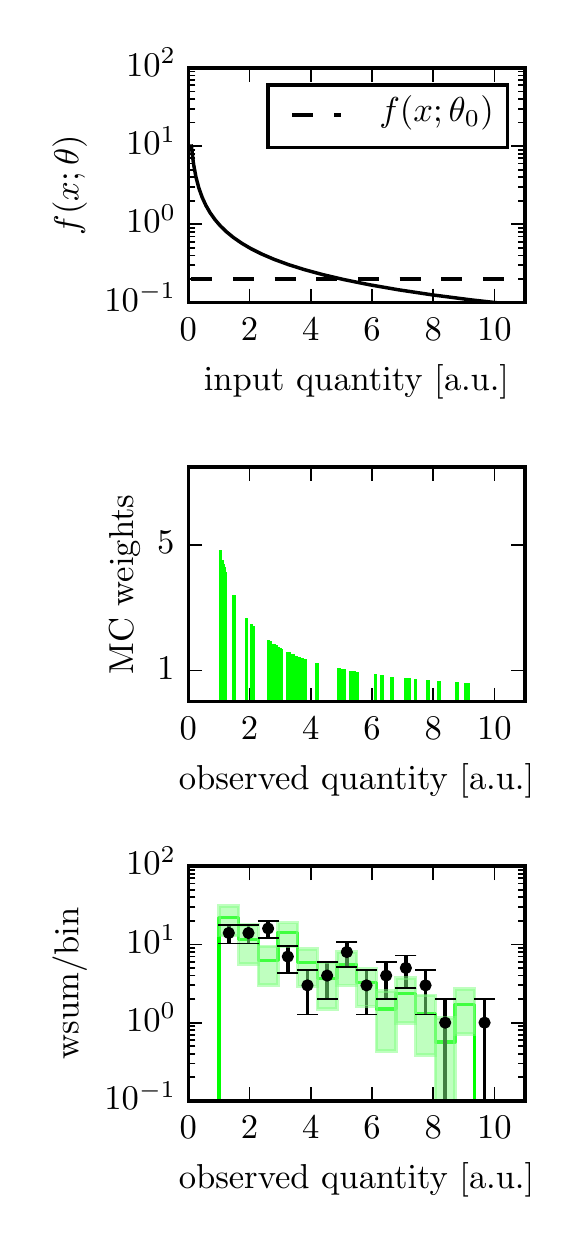}}}
\subfloat[Two different simulations for the same parameters $\theta$]{{\includegraphics{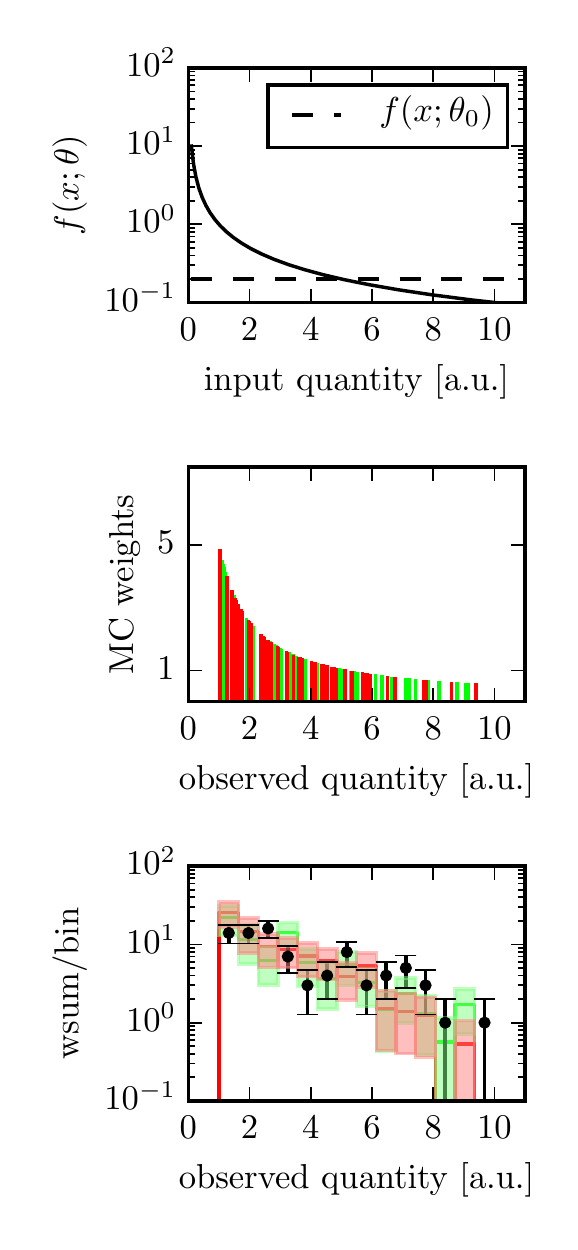}}}
\caption{Illustration of the Monte Carlo process for weighted simulation: generation function and re-weighting function (upper panel), resulting sampled and weighted MC events in the unobservable "true" dimension (central panel) and the resulting sum of MC events and some data events (black) in bins of an observable (lower panel). The uncertainty of the MC expectation is indicated by $\sum_i w_i^2$. } \label{fig:mc_illustration}
\end{figure}

Monte Carlo (\textbf{MC}) simulations are used throughout science for statistical inference. In high-energy physics, modern experiments simulate physical processes with Monte Carlo generators to approximate intractable functions that map certain physical inputs $x$ to observable outputs $y$. In experiments related to astro-particle physics, for example, cosmic-ray simulations with MC-event generators like \textit{CORSIKA} \cite{Heck1998} generate MC events following a certain cosmic-ray primary flux model with a specific composition and energy spectrum. This flux model defines the generation function $f_\mathrm{gen}(x;\theta_0)$ which depends on spectral and composition parameters $\theta_0$ and is used to draw MC events from. The input $x$ here could for example be the true primary energy $E_{\mathrm{prim}}$ of a given cosmic-ray nucleus. After a full simulation involving physical interactions in the atmosphere and the experimental detector response, the end products of the shower development are usually recorded in bins of an observable quantity by counting. In the high-energy neutrino detector IceCube \cite{Halzen2010}, such an observable quantity is for example the total charge deposited in all  Photo-multipliers from cosmic-ray induced muons \cite{Aartsen2016a} or neutrinos \cite{Aartsen2016}.  
Initially, all recorded MC events have equal weights, which is illustrated in fig. \ref{fig:mc_illustration} (a) in the central panel.
In order to save computing power, a MC simulation is often re-weighted with a function $f_{x;\mathrm{new}}(\theta)$, which might yield individual event weights different from unity (fig. \ref{fig:mc_illustration} b) given by $w_i=\frac{f_{\mathrm{new}}(x;\theta)}{f_{\mathrm{gen}}(x;\theta_0)}$. One can then bin the MC samples of the observable quantity and obtain an approximation for the expectation value in each bin $i$ as $\lambda_i=\sum_j w_j(\theta)$ (\ref{fig:mc_illustration} (a,b) in the lower panel). The MC events therefore serve as a mapping from parameters $\theta$ to expectation values $\lambda_i$, which can be used to write down a Poisson likelihood function for i.i.d. (independent and identically distributed) data as
\begin{align}
 L(\theta) = \prod_{\mathrm{bins} \ i} p(k;\theta)=\prod_{\mathrm{bins} \ i} \frac{e^{-\lambda_i} \cdot \lambda_i^{k}}{k!} \ \mathrm{with} \ \lambda_i =\sum\limits_{j=1}
 ^{N} w_j(\theta) \label{eq:poisson_weight_connection}
\end{align}. As can be seen in fig. \ref{fig:mc_illustration} (a,b) lower panel, a suitable change in $\theta$ results in a better match to the data which can be achieved via maximum-likelihood optimization \cite{Fisher1990} of $L$ (eq. \ref{eq:poisson_weight_connection}) over $\theta$.
It can also happen that re-weighting is not possible, and one has to simulate a whole new set of simulations for a given set of parameters $\theta$. This happens for example in the IceCube experiment for a specific algorithm to reconstruct neutrino directions and energies \cite{Chirkin2013} where the MC events now represent simulated Cerenkov photons emitted by a process parametrized by $\theta$.
In addition to the extra computational cost in such a scheme, each MC sample is different from a previous one, even for similar generation parameters $\theta$ (see fig. \ref{fig:mc_illustration} c). This generates extra complications for minimization routines due to fluctuations in the resulting likelihood function.

The issues we discuss in this paper arise because the Monte Carlo event count is finite which induces a statistical uncertainty in the evaluation of the related likelihood functions which depend on the MC events and their distribution in the observable bins (see eq. \ref{eq:poisson_weight_connection}). We focus on Poisson-type likelihoods of the form in (eq. \ref{eq:poisson_weight_connection}). The investigation is a continuation of a previous paper \cite{Gluesenkamp2018} which also discussed the more involved multinomial likelihood for unbinned approximations which we omit here.

In the first section, we review previous Frequentist and probabilistic approaches for this problem from the last 25 years and indicate how they are related.
In the second section, we introduce new generalized Poisson-gamma mixture distributions that model the underlying statistical more manifestly or take care of extra uncertainty so far neglected. In the third section, we discuss techniques to incorporate extra prior information into the likelihood and a scheme to incorporate empty bins. In the final section, we compare different methods in various benchmark scenarios. All results of this paper are motivated by the use case in high-energy physics, but they are generally applicable in situations where MC simulations are used to calculate Poisson likelihoods. 

\section{Previous likelihood approaches for limited Monte Carlo}

In the following we will revisit the common approaches in the literature that deal with limited Monte Carlo statistics. We will change the notation of some of them from their original publications in order to get a unified understanding. 

\subsection{Statistics of weighted Monte Carlo}
\begin{figure}
	\centering
	\includegraphics{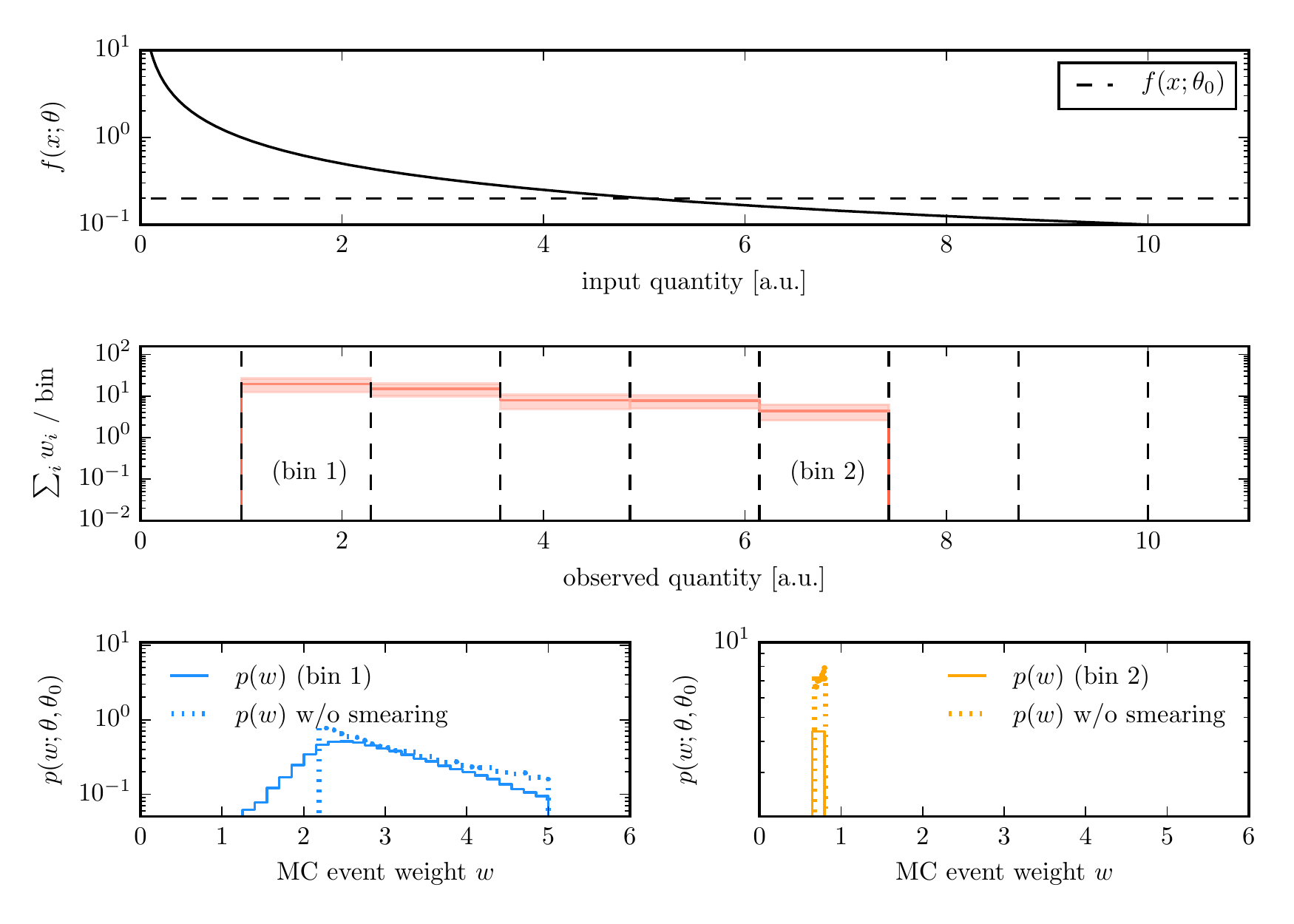}
	\caption{Illustration of the intractable true weight distribution $p(w;\theta,\theta_0)$ of the weight random variable $W$ for given parameters $\theta_0$ and $\theta$ in two example bins. The PDF is generated from many MC realizations, and shown with and without reconstruction uncertainty in the observable. The central row shows one particular realization of the sum of weights.} \label{fig:weight_stats_visu}
\end{figure}

In order to better understand the approaches in the literature, let us first take a look at the actual random variable describing weighted events in a bin $i$. For a given set of generation parameters $\theta_0$ and re-weighting parameters $\theta$, the weight of a Monte Carlo event ending up in bin $i$ follows a continuous and in general intractable distribution with random variable $W_i(\theta_0,\theta)$, where $\theta_0$ includes a parameter $N_{tot}$ that specifies the total number of events simulated \footnote{$N_{tot}$ could also be itself Poisson distributed, which would result in every $N_i$ be Poisson distributed even without the limit of the binomial distribution.}. For a given $N_{tot}$, there is a resulting Poisson distributed number of MC events $N_i$ for bin $i$ with unknown underlying mean $\mu_i=N_{tot}\cdot p$. The probability $p$ is some unknown probability for an event to land in the specific bin $i$. As long as the number of bins is sufficiently large, say $n_{\mathrm{bins}}>10$, and $p$ sufficiently small, this follows from the usual Poisson limit of the binomial distribution. The actual observed weights are samples from the PDF corresponding to $W_i$. While the underlying PDF of $W_i$ is unknown, one could imagine repeating the MC experiment with the same parameters $\theta_0$ and $\theta$ a large number of times  in order to visualize the PDF corresponding to $W_i$. Figure \ref{fig:weight_stats_visu} shows such a PDF of $W_i$ for two different bins. The PDF is shown with and without detector-related smearing in the observable, which could for example be a reconstructed energy. This illustrates that the intractable weight distribution not only depends on the parameters $\theta$ and the corresponding bin, but also on detector effects.

\begin{figure}
	\centering
	\subfloat[Equal weights]{{\includegraphics{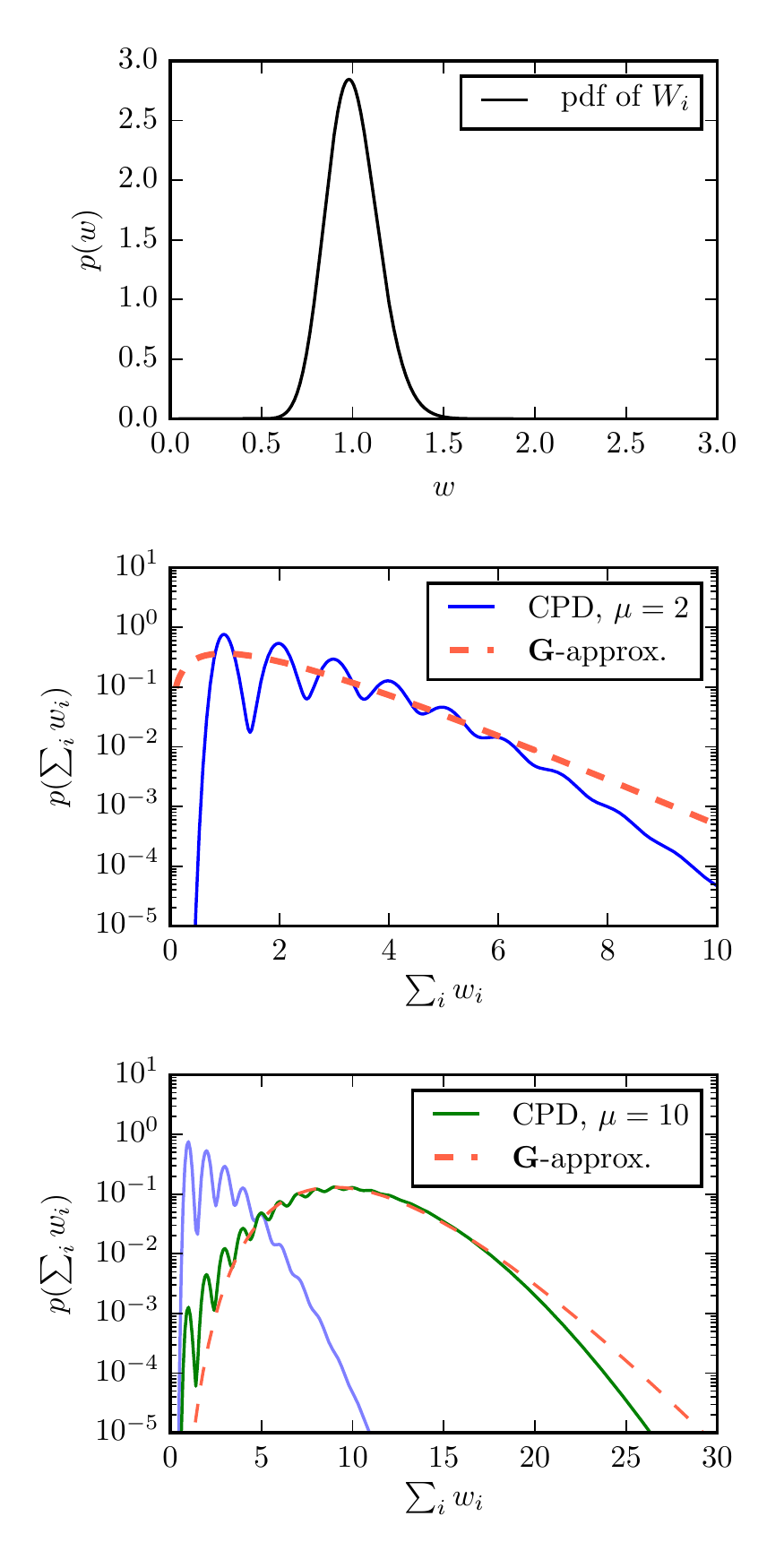}}}
	\subfloat[Re-weighted events]{{\includegraphics{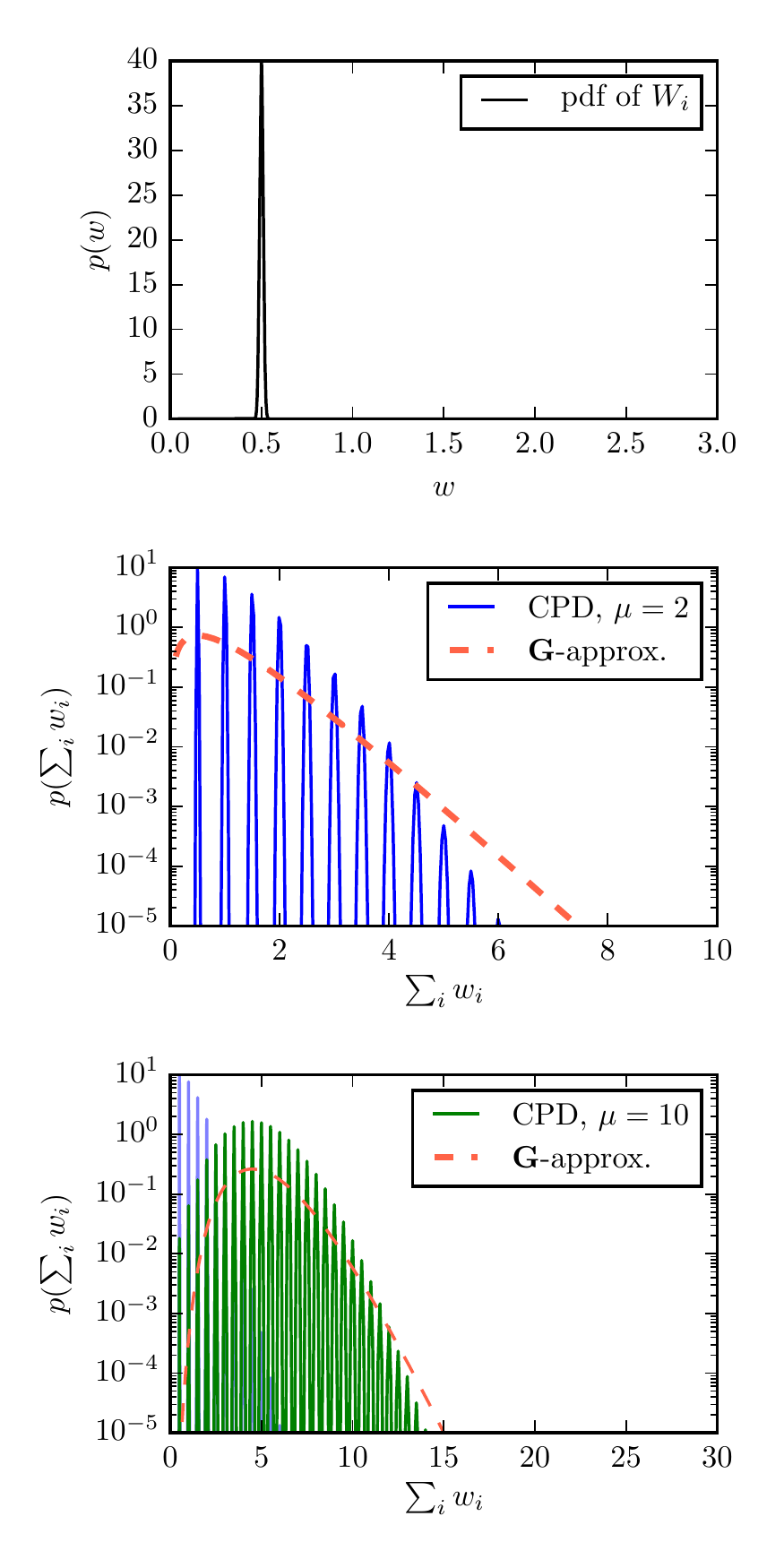}}}
	\caption{Illustration of the compound Poisson  distribution for a weight distribution with a larger spread (a) and smaller spread modeling nearly equal weights (b). The upper row shows the weight distribution $p(w)$, the central row the CPD $p(\sum_i w_i)$ for a mean of $\mu_N=2$, and the lower row for a larger mean of $\mu_N=10$. Also illustrated are approximations of the CPD distributions with gamma distributions (red dashed) that encode mean and variance of the CPD as described in eq. \ref{eq:z_mean} and eq. \ref{eq:z_variance}.} \label{fig:illustrate_cpd}
\end{figure}

From eq. \ref{eq:poisson_weight_connection} we see that we are really interested in a random variable describing the sum of weights $Z_i$ with $Z_i = \sum_{j=1}^{N_i} W_{i,j}$. From the preceding discussion we also see that $N_i$ is not fixed but itself Poisson distributed with unknown mean $\mu_i$. The resulting random variable $Z_i$ is called a compound Poisson distribution (CPD). To our knowledge this explicit statistical description of the problem was first pointed out in \cite{Bohm2014}. The problematic issue is that not only the PDF of $W_{i}$ is unknown (see fig. \ref{fig:weight_stats_visu}), but also the value of the underlying mean $\mu_i$. To get a better understanding of the CPD, figure \ref{fig:illustrate_cpd} shows two example CPDs and how they are formed from the underlying weight distribution.
All approaches described later in this section can be interpreted as different ways to approximate the CPD given the observed MC samples. 
 
The mean and variance of a compounding Poisson random variable $Z$ is given by
\begin{align}
\mathrm{E}[Z]&=\mu_z=\mathrm{E}[N] \cdot \mathrm{E}[W]  = \mu_N \cdot \mathrm{E}[W] \label{eq:z_mean}   \\
\mathrm{Var}[Z]&=\mathrm{E}[N] \cdot \mathrm{Var}[W] + (\mathrm{E}[W])^2 \cdot  \mathrm{Var}[N]\\ &\stackrel[{ \mathrm{Var}[N]=\mathrm{E}[N]} ]{}{=} \mu_N \cdot (\mathrm{Var}[W] + (\mathrm{E}[W])^2) \label{eq:z_variance}
\end{align}
where $Z=\sum_j W_j$,  $W_j$ are independent and identical realizations of a continuous random variable $W$ and $N$ is a Poisson random variable defined on  $\mathbb{N}^{+}_0$. Relationships \ref{eq:z_mean} and \ref{eq:z_variance} follow from the law of total expectation and total variance \cite{Weiss2006}, and in the last step we can exploit that mean and variance for a Poisson distribution are the same $\mu_N$. We can estimate the mean and variance using the sample mean $\widehat{\mu}_Z$ and sample variance $\widehat{\mathrm{Var}}_Z$ as
\begin{align}
\widehat{\mu}_Z&=\widehat{\mu}_N \cdot \widehat{\mu}_W   =\widehat{\mu}_N \cdot \left(\frac{1}{N}\sum_j w_j \right) = \sum_j w_j \label{eq:sample_mean} \\ 
\widehat{\mathrm{Var}}_Z&=\widehat{\mu}_N \cdot \widehat{\mathrm{Var}}_W + (\widehat{\mu}_W)^2 \cdot  \widehat{\mathrm{Var}}_N \\ &\stackrel[{ \widehat{\mathrm{Var}}_N=\widehat{\mu}_N} ]{}{=} \widehat{\mu}_N \cdot \left(\frac{1}{N}\sum_j (w_j-\widehat{\mu}_W)^2 + \widehat{\mu}_W^2\right) \\
&= \widehat{\mu}_N \cdot \left( \frac{1}{N}\sum_j w_j^2\right) 
=\sum_j w_j^2 \label{eq:sample_var} 
\end{align}
where we just replace every term in eqs. \ref{eq:z_mean} and \ref{eq:z_variance} by the respective sample estimate, using $w_j$ to denote
the observed weights. 
We used the biased sample variance of weights, but could have used the unbiased variance estimator instead. 
The resulting estimates are incidentally also the sample mean and sample variance of the sum of $N$ weighted Poisson random variables with $\lambda=1$ each, and they are widely used for plotting purposes to indicate the uncertainty of weighted MC histograms. The mean $\widehat{\mu}_N$ is unknown and by approximating it with $N$ there is always some sample uncertainty neglected. In the following we will sometimes use $N$ to denote number of Monte Carlo events or sometimes explicitly call it $k_{mc}$.

\subsection{Frequentist Approaches}
\label{sec:frequentist_approaches}
The approaches in this section are called "Frequentist" because they can be thought of as "profile likelihoods" \cite{Murphy2000} with additional nuisance parameters.

\subsubsection{Barlow/Beeston (1993)}
The systematic treatment of limited MC statistics for Poisson likelihood applications goes back at least 25 years to Barlow and Beeston \cite{Barlow1993}. The authors augment the Poisson likelihood with additional Poisson factors for MC datasets, effectively treating MC as additional data. Instead of a per-bin likelihood as described in eq. \ref{eq:poisson_weight_connection}, the modified likelihood per bin looks like
\begin{align}
 L_{\mathrm{bin}} &= \frac{e^{-\sum_j p_j \widehat{w}_j \lambda_j} \cdot (\sum_j p_j \widehat{w}_j \lambda_j)^k}{k!}  \cdot \prod_j^{N_{\mathrm{src}}} \frac{e^{-\lambda_j} \cdot (\lambda_j)^{k_{mc,j}}}{k_{mc,j}!}  \label{eq:barlow_1}
\end{align}
Each factor $j$ enumerates $N_{src}$ individual MC sources\footnote{We use slightly different notation from the original publication \cite{Barlow1993}. We write $k_{mc,j}$ instead of $a_{ji}$, $\lambda_j$ instead of $A_{ji}$, and directly use weights as described in section 6 in \cite{Barlow1993}.}, for example a signal and a background dataset. The weights of Monte Carlo events are averaged per dataset as $\widehat{w}_j$. The parameters $\lambda_j$ are nuisance parameters and not of physical interest, and they should be optimized to maximize the values of the individual Poisson factors. The $p_j$ are global all-bin strength factors and can also be optimized as parameters, or fixed if relative strengths are known - in fact they could be incorporated into the weights directly. They are shared for all bins, and an overall scaling by $p_j \rightarrow c\cdot p_j$ of source $j$ would just increase the weights of source $j$ in all bins by the scaling factor $c$ in this case, making $p_j$ redundant if the weights are trusted in an absolute manner. This is actually the case in many modern applications. The nuisance optimization itself effectively handles some of the uncertainty from the finite MC event count. The authors \cite{Barlow1993}  further show how the $j$ parameters $\lambda_j$ can be reduced to a single one per bin by taking the logarithm and finding the extremal values of $\mathrm{ln}(L_{\mathrm{bin}})$ in $\lambda_j$, which results in coupled equations for the $\lambda_j$ where $j-1$ of them can be eliminated. The resulting important relations eq. (25) and eq. (26) from  \cite{Barlow1993} look like 
\begin{align}
\frac{k}{(1-t)} = \sum_j \lambda_j \cdot p_j \widehat{w}_j = \sum_j \frac{k_{mc,j}\cdot p_j \widehat{w}_j}{1+p_j \widehat{w}_j \cdot t}  \label{eq:barlow_relations}
\end{align}
in our notation.
Taking the logarithm of eq. \ref{eq:barlow_1}, we can write 
\begin{align}
\mathrm{ln}(L_{\mathrm{bin}})&=-\sum_j p_j \widehat{w}_j \lambda_j+k\cdot \mathrm{ln}(\sum_j p_j \widehat{w}_j  \lambda_j) -\mathrm{ln}(k!)  + \sum_j \big[-\lambda_j + k_{mc,j} \cdot \mathrm{ln}(\lambda_j) - \mathrm{ln}(k_{mc,j}!)\big] \\
&=-\sum_j \frac{(1+p_j \widehat{w}_j)\cdot k_{mc,j} }{1+p_j \widehat{w}_j \cdot t} + k \cdot \mathrm{ln}(
\frac{k}{1-t}) + \sum_j k_{mc,j} \cdot \mathrm{ln}(\frac{  k_{mc,j}}{1+p_j\widehat{w}_j \cdot t})  - \sum_j \mathrm{ln}(k_{mc,j}!) - \mathrm{ln}(k!) \label{eq:barlow_2nd} \\
&= - k - k_{mc,\mathrm{tot}} + k \cdot \mathrm{ln}(\frac{k}{1-t}) + \sum_j k_{mc,j} \cdot \mathrm{ln}(\frac{  k_{mc,j}}{1+p_j\widehat{w}_j \cdot t})  - \sum_j \mathrm{ln}(k_{mc,j}!) - \mathrm{ln}(k!) \label{eq:barlow_3rd}
\end{align}
 where we use eq. \ref{eq:barlow_relations} and insert $\lambda_j=\frac{k_{mc,j}}{1+p_j\widehat{w}_j \cdot t}$ to reduce the $j$ parameters $\lambda_j$ to a single parameter $t$ except for the first logarithm where we replace $p_j\widehat{w}_j \cdot \lambda_j$ with $\frac{k}{1-t}$. We further simplified eq. \ref{eq:barlow_2nd} to eq. \ref{eq:barlow_3rd} using the reformulation
\begin{align}
\sum_j \frac{(1+p_j \widehat{w}_j)\cdot k_{mc,j} }{1+p_j \widehat{w}_j \cdot t}&=\sum_j \frac{(1+p_j \widehat{w}_j \cdot t - p_j \widehat{w}_j \cdot t + p_j \widehat{w}_j)\cdot k_{mc,j} }{1+p_j \widehat{w}_j \cdot t}\\
&=k_{mc,\mathrm{tot}}+\sum_j \frac{p_j \widehat{w}_j \cdot (1-t)\cdot k_{mc,j} }{1+p_j \widehat{w}_j \cdot t} = k_{mc,\mathrm{tot}} + k
\end{align} 
again making use of eq. \ref{eq:barlow_relations}. Assuming the $p_j$ are fixed, for example when they can be absorbed into the average weights $\widehat{w}_j$, we only have to find the value of $t$ that maximizes expression eq. \ref{eq:barlow_3rd} for every bin. We can form the derivative of eq. \ref{eq:barlow_3rd} with respect to $t$ and obtain 
\begin{align}
\frac{d}{dt} \mathrm{ln}(L_{\mathrm{bin}})=0=\frac{k}{(1-t)} - \sum_j \frac{k_{mc,j}\cdot p_j \widehat{w}_j}{1+p_j \widehat{w}_j \cdot t} 
\end{align} which is just reproducing eq. \ref{eq:barlow_relations}. When there is only one source dataset this equation can be solved for $t$ exactly and one obtains
\begin{equation}
t=\frac{k_{mc,1}\cdot p_1 \widehat{w_1}-k}{ p_1 \widehat{w_1} \cdot (k+k_{mc,1})}
\end{equation} which using eq. \ref{eq:barlow_relations} can be used to substitute $\lambda_1$ in eq. \ref{eq:barlow_1} which yields the solution
\begin{align}
L_{\mathrm{bin},\mathrm{eq.}}= \frac{e^{-(k+k_{mc,1})} }{k! k_{mc,1}!} \cdot \left(\frac{k+k_{mc,1}}{1+1/(p_1 \widehat{w}_1)}\right)^{k} \cdot \left(\frac{k+k_{mc,1}}{1+p_1 \widehat{w}_1}\right)^{k_{mc,1}}
\label{eq:barlow_equal_weights}
\end{align} with no free parameters. This is also the exact solution when all weights are equal and the average weight $\widehat{w}_1$ is actually not averaged. In practice, one would insert $t$ into the log-likelihood expression eq. \ref{eq:barlow_3rd}, but it is instructive to compare eq. \ref{eq:barlow_equal_weights} with other methods later. For more than one dataset we have to find the optimal value of $t$ numerically. The authors also discuss the case when no MC events are present in a bin. In that case they argue that $t$ should only depend on the largest $p_j$, which then can be used again to calculate the $\lambda_j$. While the method by Barlow/Beeston can not be directly interpreted as an approximation of the overall CPD, it can be interpreted as approximating individual CPD's for each dataset, whose random variables are then summed. This will be exploited later in section \ref{sec:further_generalizations} for a probabilistic generalization.

\subsubsection{Chirkin (2013)}

In 2013 Chirkin \cite{Chirkin2013b} implicitly derived a certain generalization of eq. \ref{eq:barlow_1}. The author starts with a multinomial likelihood and derives a formula that looks very similar to eq. \ref{eq:barlow_relations} in the setting of what he calls "weighted simulation without model errors". In fact, the resulting equation is the same if we absorb the relative fractions $p_j$ into the weights and if we then treat each MC event independently as coming from its own "source", i.e. $k_{mc,j}=1$. From eq. \ref{eq:barlow_relations} we then obtain 
\begin{align}
\frac{k}{(1-t)} =\sum\limits_{i=1}^{k_{mc,tot}} \frac{w_i}{1+w_i \cdot t}
\end{align} which is similar to the equation obtained in section (5) in \cite{Chirkin2013b}\footnote{Using the notation of \cite{Chirkin2013b} we have $d_k=k$, $t_{ki}=1/w_i$, $t_d=1$, $\xi_k=t$, $s_{ki}=1$.}. The resulting corresponding likelihood formulation equivalent to eq. \ref{eq:barlow_1} is 
\begin{align}
L_{\mathrm{bin}}= \frac{e^{-\sum_i w_i \lambda_i} \cdot (\sum_i  w_i \lambda_i)^k}{k!}  \cdot \prod_{i=0}^{k_{mc,tot}} \frac{e^{-\lambda_i} \cdot (\lambda_i)^{1}}{1!}  \label{eq:chirkin}
\end{align} which now has a product over all individual MC events instead of whole datasets with averaged weights. The method by Chirkin can therefore be interpreted as a certain generalization of the Barlow/Beeston method. It omits weight averaging and gives better results for bias reduction in likelihood scans for a single weighted dataset as demonstrated in \cite{Gluesenkamp2018}, and sometimes also for multiple datasets (see later comparison). If all weights are equal the two methods are similar and their likelihoods are given by eq. \ref{eq:barlow_equal_weights} up to constant factors. The method can also be extended to include systematic uncertainties in a log-normal term that is dubbed "model error" or handle bins with no MC using a constant "noise" term. These terms have to be added by hand and are somewhat arbitrary.

\subsubsection{Bohm/Zech (2012)}

Bohm and Zech discuss another possibility to set up a pseudo-likelihood via
\begin{align}
 L_{\mathrm{bin}}= \frac{e^{-\lambda} \cdot \lambda^k}{k!} \cdot \frac{e^{-\lambda \cdot \beta  } \cdot (\lambda\cdot \beta)^{\alpha} }{\Gamma(\alpha+1)}
\label{eq:bohm_zech_pseudollh}
\end{align} where $\lambda$ is a nuisance parameter, $\alpha=\frac{(\sum_i w_i)^2}{\sum_i w_i^2}$ and 
$\beta=\frac{\sum_i w_i}{\sum_i w_i^2}$. A single Poisson factor is used to encode the mean and variance of the sum of weighted MC in terms of effective counts to approximate the CPD. Because of the gamma function instead of the factorial this is not a real Poisson  distribution and therefore not a real likelihood, so in general it can not be used for likelihood applications. The authors also only discuss it in the context of goodness of fit tests where they argue it is sensible. For equal weights, however, the gamma function turns into a factorial and the expression is similar to the previous two approaches and eq. \ref{eq:bohm_zech_pseudollh} turns into eq.  \ref{eq:barlow_equal_weights}. It will still be interesting to compare the general pseudo-likelihood (eq. \ref{eq:bohm_zech_pseudollh}) to its probabilistic counterpart discussed later.

\subsection{Probabilistic approaches}

In contrast to the previous Frequentist approaches this section deals with solutions that integrate nuisance parameters instead of optimizing them. 

\subsubsection{Gl\"usenkamp (2018)}

In a previous paper \cite{Gluesenkamp2018} we discussed a generalization of the Poisson likelihood of the form
\begin{align}
L_{bin}(\theta)&= \frac{e^{-\sum_i w_i} \cdot (\sum_i w_i)^k}{k!} = \int_{0}^{\infty} \frac{{e^{-\lambda}}{\lambda}^{k} }{k!} \cdot \delta(\lambda - \sum_i w_i) d \lambda \\ &=
\int_{0}^{\infty} \frac{{e^{-\lambda}}{\lambda}^{k} }{k!} \cdot \left[\delta(\lambda-w_1) \ast \ldots \ast \delta(\lambda-w_N)\right](\lambda) \ d\lambda
\\ \rightarrow L_{bin}(\theta) &= \int_{0}^{\infty} \frac{{e^{-\lambda}}{\lambda}^{k} }{k!} \cdot \left[\mathrm{\mathbf{G}}(\lambda_{1};1+\alpha/N,1/w_{1}) \ast \ldots \ast \mathrm{\mathbf{G}}(\lambda_{N};1+\alpha/N,1/w_{N})\right](\lambda) \ d\lambda
\label{eq:prob_last_paper}
\end{align} which can be thought of as directly expanding the delta factors into gamma distributions. Each gamma distribution can be derived via Bayesian inference and approximates the underlying weight distribution $W_i$ of the CPD. This means a convolution of these gamma factors is some approximation of the CPD (see  fig \ref{fig:cpd_relationships}) , the difference being a fixed number of terms and different individual PDFs. There is some prior freedom in this approach, which we parametrized by a parameter $\alpha$ which we here call $\alpha^{`}$. The parameter is shared among individual gamma distributions such that for equal weights it reduces to a single gamma factor with shape parameter $N+\alpha^{`}$. In \cite{Gluesenkamp2018} we observed optimal performance in terms of likelihood ratio bias with respect to infinite statistics for $\alpha^{`}=0$ for a single weighted dataset. We will later see that this is not true in general. The setting $\alpha=0$ yields the expected first and second of the CPD as $ \mu_{G_1 \ast \ldots \ast G_N}=\sum_i w_i$ and $ \mathrm{var}_{G_1 \ast \ldots \ast G_N}=\sum_i w_i^2$. It also showed practically identical likelihood ratio behavior as the method by Chirkin \cite{Chirkin2013b} for a single source dataset\cite{Gluesenkamp2018}. This is not surprising since a comparison of eq. \ref{eq:prob_last_paper} with eq. \ref{eq:chirkin} shows that it can be interpreted as the probabilistic counterpart to the approach by Chirkin \cite{Chirkin2013b} which enforces the sum constraint via optimization instead of probabilistic convolution (see fig. \ref{fig:cpd_relationships}). The solution of eq. \ref{eq:prob_last_paper} was derived in \cite{Gluesenkamp2018} as an iterative sum via
\begin{align}
L_{bin}(\theta) = D_k \cdot \prod\limits_{i=1}^{N} \left(\frac{1}{1+w_i}\right)^{1+\alpha/N} \label{eq:poisson_gamma_mixture_simple_iterative}
\end{align}
where
\begin{align}
D_k=\frac{1}{k}\sum\limits_{j=1}^{k} \left[\left(\sum\limits_{i=1}^{N} (1+\alpha/N) \cdot \left({\frac{1}{1+1/w_i}}\right)^j \right) D_{k-j}\right]
\end{align} with $D_0=1$. The formula allows for an efficient computation \footnote{See also appendix \ref{appendix:general_formulas} for the general form and different forms of writing}. For equal weights, the result can be written as a simple Poisson gamma mixture as 
\begin{align}
L_{\mathrm{bin, eq.}} &=
\mathrm{E}\left[\frac{{e^{-\lambda}}{\lambda}^{k} }{k!}\right]_{G(\lambda; N+\alpha, 1/w)} \nonumber \\
& = \frac{(1/w)^{N+\alpha}  \cdot \Gamma(k+N+\alpha)}{  \Gamma(N+\alpha) \cdot k!\cdot(1+1/w)^{k+N+\alpha}}  \label{eq:prob_last_paper_equal_weights}
\end{align}, where again we use $N=k_{mc}$. Comparing eq. \ref{eq:prob_last_paper_equal_weights} with eq. \ref{eq:barlow_equal_weights} we see that for equal weights the approach has similar dependence on the weights $w$ as the Frequentist approaches using the unique prior $\alpha=0$. Likelihood scans with equal weights therefore give numerically identical confidence intervals as the Frequentist methods.

\subsubsection{Arg\"uelles et al. (2019)}
\label{sec:say_llh}
\begin{figure}
\begin{tikzpicture}[node distance=1cm, auto,]
\node[box] (chirkin) {\centering Chirkin (2013) \cite{Chirkin2013b} \\
	$ \displaystyle \max_{\{\bm{\lambda}\}} \mathrm{\mathbf{P}}(k;\sum_i  w_i \lambda_i)  \cdot \prod_i^{N} \mathrm{\mathbf{P}}(1;\lambda_i)$ \\
	or \\
$\displaystyle \max_{\{\bm{\lambda^*}\}} \mathrm{\mathbf{P}}(k;\sum_i  {\lambda_i}^*)  \cdot \prod_i^{N} \mathrm{\mathbf{G}}({\lambda_i}^*;2,\frac{1}{w_i}) \cdot w_i$};

\node[box, above=of chirkin] (barlow) {\centering Barlow/Beeston (1993) \cite{Barlow1993}: \\
	$\displaystyle \max_{\{\bm{\lambda}\}} \mathrm{\mathbf{P}}(k;\sum_j p_j \widehat{w}_j \lambda_j)  \cdot \prod_j^{N_{\mathrm{src}}} \mathrm{\mathbf{P}}(k_{mc,j};\lambda_j)$ \\
	or \\
$\displaystyle \max_{\{\bm{\lambda^*}\}} \mathrm{\mathbf{P}}(k;  {\lambda_j}^*)  \cdot \prod_j^{N_{\mathrm{src}}} \mathrm{\mathbf{G}}({\lambda_j}^*; k_{mc,j}+1;\frac{1}{p_j \widehat{w}_j}) \cdot p_j \widehat{w}_j$}
edge[fw_arrow, bend left=12] node[swap, near end] {$N_{\mathrm{src}} = N$, absorb $p_j$} (chirkin.north);

\node[box, above=of barlow] (bohm_zech) {\centering Bohm/Zech (2012) \cite{Bohm2012}: \\ $\displaystyle \max_{\{\lambda\}} \mathrm{\mathbf{P}}(k;\lambda) \cdot \frac{e^{-\lambda \cdot \beta  } \cdot (\lambda\cdot \beta)^{\alpha} }{\Gamma(\alpha+1)}$ \\ $\alpha=\frac{(\sum_i w_i)^2}{\sum_i w_i^2}$ , 
	$\beta=\frac{\sum_i w_i}{\sum_i w_i^2}$};
\node[solidbox, above=of bohm_zech] (header_left) {Frequentist};

\node[box,  right=3 cm of chirkin] (pg1) {\centering Gl\"usenkamp (2018) \cite{Gluesenkamp2018} \\ $\displaystyle \int \mathrm{\mathbf{P}}(k;\lambda) \cdot \left[\mathrm{\mathbf{G}}(\lambda_1;1+\frac{\alpha}{N},\frac{1}{w_1})\ast \ldots \ast \mathrm{\mathbf{G}}(\lambda_{N};1+\frac{\alpha}{N},\frac{1}{w_{N}}) \right](\lambda) d \lambda$}
edge[bw_arrow, bend left=12] node[auto] {Prob. counterpart} (chirkin);

\node[solidbox, right=2.5 cm of barlow] (cpd_stats) {\centering 
$Z=\sum_{i=1}^{N} W_i$ \cite{Bohm2014} \\ $N\sim \mathrm{Poisson}$ \\
$\widehat{\mu}(Z)=\sum_i w_i$ \\
$\widehat{\mathrm{var}}(Z)=\sum_i w_i^2$ \\}
edge[fw_arrow, bend right=12]  node[pos=0.6] {encode $\widehat{\mu}(Z)$ and $\widehat{\mathrm{var}}(Z)$ } (bohm_zech)
edge[fw_arrow, bend left=12] node[near end] {encode sum of $W_i$} (chirkin)
edge[fw_arrow, bend right=12] node[auto] {approximate $Z$, $\mathrm{\mathbf{G}} \approx W_i$, $N$ fixed} (pg1);
\node[solidbox, above=0.1cm of cpd_stats] {Statistics of\\ weighted MC};
\node[box, right=7cm of bohm_zech] (say) {\centering Arg\"uelles et al. (2019) \cite{Argueelles2019} \\ $\displaystyle \int \mathrm{\mathbf{P}}(k;\lambda) \cdot  \mathrm{\mathbf{G}}(\lambda;\alpha,\beta) d\lambda$ \\ $\alpha=\frac{(\sum_i w_i)^2}{\sum_i w_i^2}+a$ , $\beta=\frac{\sum_i w_i}{\sum_i w_i^2}+b$}
edge[bw_arrow, bend left=12] node[auto] {encode $\widehat{\mu}(Z)$ and $\widehat{\mathrm{var}}(Z)$} (cpd_stats)
edge[bw_arrow, bend right=12] node[swap] {Prob. counterpart} (bohm_zech);
\node[solidbox, above=of say] (header_right) {Probabilistic};

\node[box, left=0.9cm of cpd_stats,draw=none] (fake) {        };
\node[solidbox, below=-0.2cm of fake] (small_cpd) {\centering $Z=\sum_j Z_j$}
edge[fw_arrow, bend right=12] node[swap, align=center,pos=0.0] {encode sum \\ of $Z_j$} (barlow);

\end{tikzpicture}
\caption{Relationships between existing Frequentist and probabilistic approaches in the literature that extend the standard Poisson likelihood for a given bin. $N$ denotes the total number of MC events, $N_{src}$ the total number of source datasets, $\mathrm{\mathbf{P}}$ the Poisson distribution, $\mathrm{\mathbf{G}}$ the gamma distribution.}
\label{fig:cpd_relationships}
\end{figure}
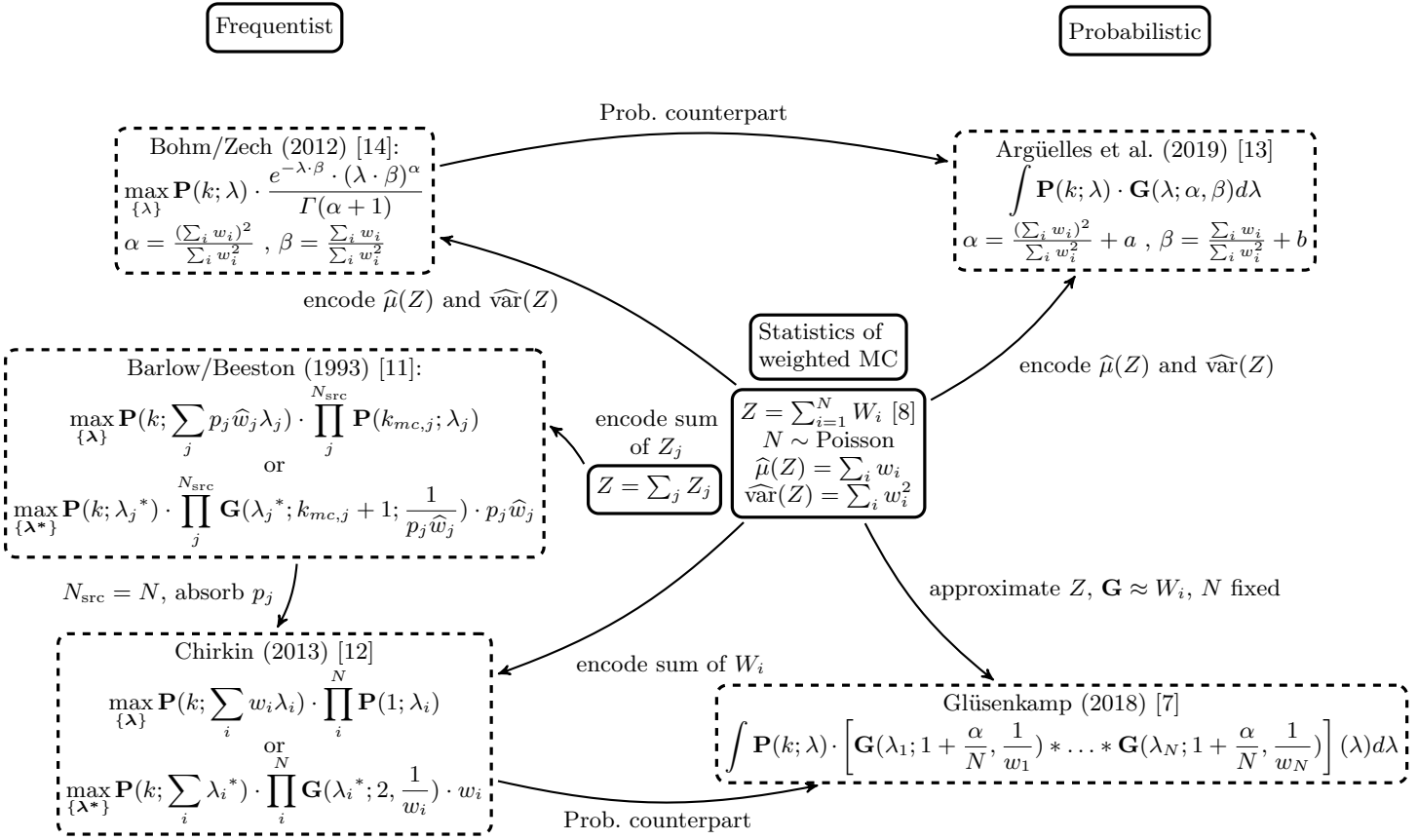

In Arg\"uelles et al. \cite{Argueelles2019} the authors describe an Ansatz to encode the mean and variance of the CPD in a single gamma distribution which effectively approximates the CPD. It looks like
\begin{align}
L_{bin}(\theta)&=\int \frac{e^{-\lambda} \cdot \lambda^k}{k!} \cdot \frac{e^{-\lambda \cdot \beta  } \cdot (\lambda\cdot \beta)^{\alpha} }{\Gamma(\alpha+1)} d\lambda
\end{align} where they add some additional effective parameters $a$ and $b$ to expand its possibilities. It can be thought of as a direct probabilistic counterpart of the pseudo-likelihood (eq. \ref{eq:bohm_zech_pseudollh}) described by Bohm and Zech \cite{Bohm2012} (see figure \ref{fig:cpd_relationships}). In contrast to the pseudo likelihood, this Ansatz involves a marginalization with a proper PDF and therefore works for arbitrary weights. The authors argue that the choice of $a=1$, $b=0$ works best in their test in terms of proper coverage, in particular better than the choice $a=0$, $b=0$ which would just encode the sample mean and sample variance of the CPD in the gamma distribution. We will see later that this is not the case in general, in particular that it depends on the number of datasets combined and the MC statistics per dataset. In section \ref{sec:generalization_b} we describe an alternative parametrization that directly parametrizes the unknown Poisson mean and unknown weight moments which we later use to incorporate extra Prior information (section \ref{sec:incorporating_prior_info}).

\subsection{Summary}

The relationship between the different approaches is highlighted in fig. \ref{fig:cpd_relationships}. It rests on the interpretation that all approaches fundamentally approximate the CPD, an interpretation that has previously only been explictly made for the formulas put forward by Bohm/Zech \cite{Bohm2012} and Arg\"uelles et al. \cite{Argueelles2019} and vaguely for our previous construction in \cite{Gluesenkamp2018}. For the Frequentist approaches by Barlow/Beeston \cite{Barlow1993} and Chirkin \cite{Chirkin2013b}, the prevailing interpretation has been that the Monte Carlo data is simply treated as additional data, as indicated in fig. \ref{fig:cpd_relationships} by the upper formulas, respectively. However, by a simple change of variables from $\lambda_i$ to $\lambda_i^*$, one can see that one obtains a formula  that almost resembles the probabilistic counterpart, in the case of Chirkin it resembles the probabilistic formula we proposed earlier in \cite{Gluesenkamp2018}. In this interpretation, the Frequentist approaches therefore also approximate the CPD via optimization in $\lambda_i$, which now resemble nuisance parameters where each parameter has its own gamma Prior. It rests on the property of the Poisson distribution that it can always be interpreted as a corresponding gamma distribution, whose shape parameter is shifted by one. It turns out that in order for an optimization to give similar results as the corresponding marginalization, the mode of the gamma distribution for optimization has to match the mean during marginalization. This is exactly fulfilled if the shape $\alpha$ is larger by one on the Frequentist side, since the mode of the gamma distribution is $\frac{\alpha-1}{\beta}$, while the mean is $\frac{\alpha}{\beta}$. The only difference remaining is the extra scaling factor $w_i$, that is being pulled out of the gamma factor. This, however, would exactly be the inverse of the Jacobian factor from the variable transform $\lambda_i^*=w_i \cdot \lambda_i$ and cancel, if one actually performed an integration.

\section{Exact CPD for equal weights}
\label{sec:exact_cpd}

What if we knew the PDF of the exact CPD, $p_{CPD}(\lambda)$, and integrate over it 
\begin{align}
L_{bin,exact}=\int \frac{e^{-\lambda} \cdot \lambda^k}{k!} \cdot p_{CPD}(\lambda) d\lambda
\end{align}
instead of using gamma factors in some form?
In the special case of equal weights we can actually do this. In \cite{Bohm2014} it was argued that the we can approximate the CPD with a scaled Poisson distribution, but for equal weights the CPD *is* a scaled Poisson distribution. This results in $p(w)=\delta(\lambda-w)$ and $p_{\mathrm{fix}}(\sum_i w_i)=p(w_1) \ast \ldots \ast p(w_N)=\delta(\lambda-k_{mc}\cdot w)$, which yields 
\begin{align}
p_{CPD}(\lambda) = \sum\limits_{k_{mc}=0}^{\infty} \mathrm{\mathbf{P}}(k;\lambda) \cdot p_{\mathrm{fix}}(\lambda) =  \sum\limits_{k_{mc}=0}^{\infty} \frac{e^{-\mu} \cdot \mu^{k_{mc}}}{k_{mc}!} \cdot \delta(\lambda-k_{mc}\cdot w)
\end{align} which is a superposition of the Poisson PDF to see $N$ weighted events with the probability distribution $p_{\mathrm{fix}}$ to see the weight sum of a given $N$ weighted events. So instead of approximating the CPD with gamma factors (see fig. \ref{fig:illustrate_cpd}) we can write down the exact CPD and integrate over it. The exact CPD for equal weights looks  roughly equivalent to the example in figure \ref{fig:illustrate_cpd} (b) where the weight distribution is very narrow. The final likelihood then can be written as
\begin{align}
L_{bin,exact,equal}&=\int  \frac{e^{-\lambda} \cdot \lambda^k}{k!} \cdot \sum\limits_{k_{mc}=0}^{\infty} \frac{e^{-\mu} \cdot \mu^{k_{mc}}}{k_{mc}!} \cdot \delta(\lambda-k_{mc}\cdot w) d \lambda \label{eq:cpd_exact_int} \\
&=\sum\limits_{k_{mc}=0}^{\infty} \frac{e^{-k_{mc}w} \cdot (k_{mc}w)^k}{k!} \cdot \frac{e^{-\mu} \cdot \mu^{k_{mc}}}{k_{mc}!} 
\label{eq:cpd_exact}
\end{align} which can be solved numerically. The parameter $\mu$ is an unknown mean, but can be determined by the average of many Monte Carlo runs. The result is a likelihood that does not depend on individual MC realizations anymore and involves the true PDF of the sum of weights without any approximations. It will serve as a crosscheck in the next section.

\section{Further generalizations}
\label{sec:further_generalizations}
\begin{figure}
	\begin{tikzpicture}[node distance=1cm, auto]
	\node[box] (header_left) { \centering Interpretation 1: \\
	$\mathbf{G}$ encodes weight distributions $W_i$};

\node[solidbox,  below=3 cm of header_left] (interp1) {\centering Generalization (1) \\ $ \displaystyle \int \mathrm{\mathbf{P}}(k;\lambda) \cdot \left[\mathrm{\mathbf{GPG}}_1 \ast \ldots \ast \mathrm{\mathbf{GPG}}_N \right](\lambda) d \lambda$}
edge[bw_arrow, bend left=12] node[align=center] {given Gl\"usenkamp (2018)\cite{Gluesenkamp2018},\\marginalize discrete shape of each $\mathbf{G}$ \\ with Poisson-Gamma mixture (PG)} (header_left);


\node[box, right=3 cm of header_left] (header_right) { \centering Interpretation 2: \\
	$\mathbf{G}$ encodes the total CPD};

\node[solidbox, below=3 cm of header_right] (interp2) { \centering Generalization (2)  \\
	$\displaystyle \int \mathrm{\mathbf{P}}(k;\lambda) \cdot \left[\mathrm{\mathbf{G}}_1\ast \ldots \ast \mathrm{\mathbf{G}}_{N_{src}} \right](\lambda) d \lambda$ \\
with $\mathrm{\mathbf{G}}_j=\mathrm{\mathbf{G}}(\mu_j \cdot \alpha_j, \beta_j)$, $\alpha_j=\frac{(\sum_i w_{i,j})^2}{k_{mc,j} \cdot \sum_i w_{i,j}^2}$ , $\beta_j=\frac{\sum_i w_{i,j}}{\sum_i w_{i,j}^2}$ \\ }
edge[bw_arrow, bend right=12] node[align=center] {one CPD for each dataset $j$, \\ $Z_{tot}=\sum_j Z_j$, \\ each $\mathbf{G}$ encodes its own CPD, \\ multiplicative parametrization via $\mu$} (header_right);

\node[solidbox, below=2 cm of interp2] (interp3) { \centering Generalization (3)  \\
	$\displaystyle \int \mathrm{\mathbf{P}}(k;\lambda) \cdot \left[\mathrm{\mathbf{GG}}_1\ast \ldots \ast \mathrm{\mathbf{GG}}_{N_{src}} \right](\lambda) d \lambda$ \\
	with $\mathrm{\mathbf{GG}}_j=\mathrm{\mathbf{GG}}(\lambda_j;\alpha_j, \beta_j,k_{mc};1)$ \\ }
edge[bw_arrow, bend right=12] node[align=center] {marginalize $\mu_j$ with $\mathbf{G}(\mu;k_{mc};1)$ \\  instead of just setting $\mu=k_{mc}^*$ \\ (captures uncertainty in number of events)} (interp2);

	\end{tikzpicture}
	\caption{Further generalizations based on two different fundamental interpretations. $N$ denotes the total number of MC events, $N_{src}$ the total number of source datasets, $\mathrm{\mathbf{P}}$ the Poisson distribution, $\mathrm{\mathbf{G}}$ the gamma distribution, $\mathrm{\mathbf{GPG}}$ a gamma-Poisson-gamma mixture, and  $\mathrm{\mathbf{GG}}$ a gamma-gamma mixture distribution. The mixtures integrate out shape parameters. $Z$ denotes the random variable for the sum of weights, either for all datasets ($Z_{tot}$), or for individual individual datasets ($Z_j$).}
\label{fig:further_generalizations}
\end{figure}
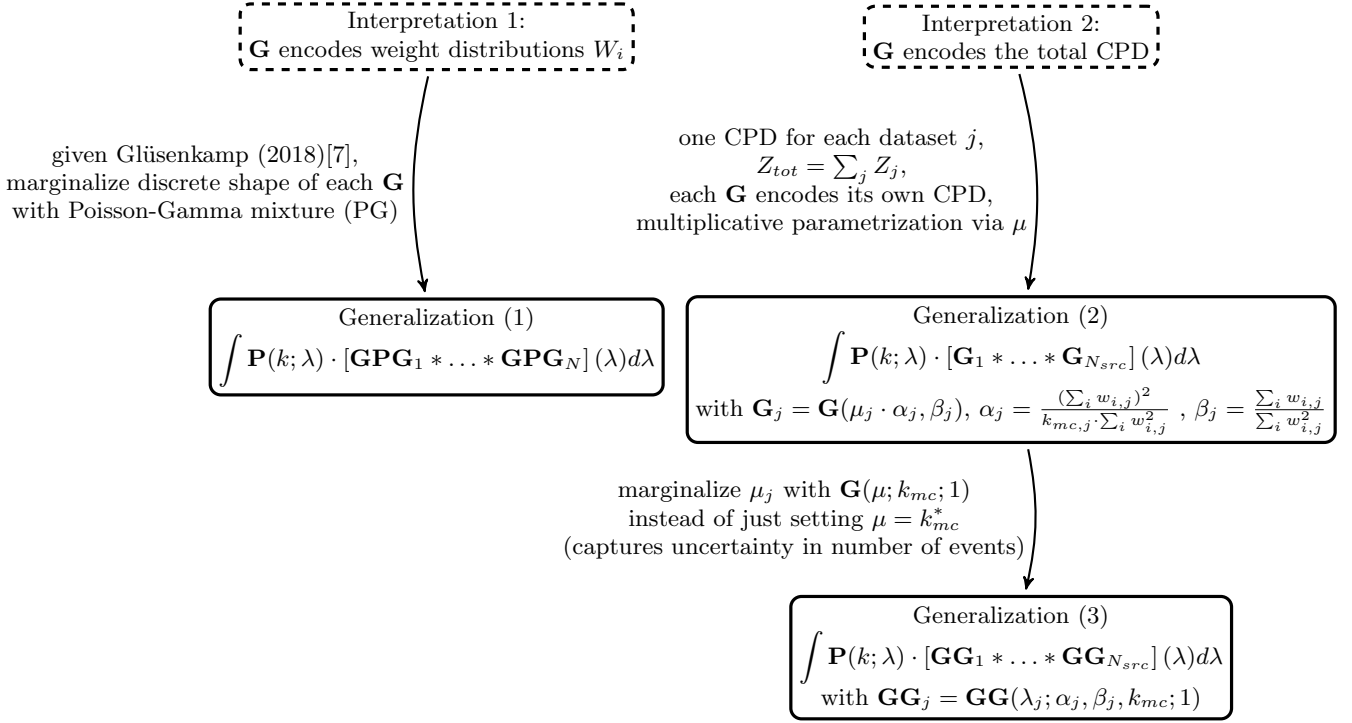

We will now discuss some further constructions which generalize the previously discussed probabilistic approaches. An overview is shown in figure \ref{fig:further_generalizations}, which we will discuss in detail in the following.

\subsection{Generalization 1 - discrete marginalization of $k_{mc}$ with a Poisson-gamma mixture}
\label{sec:generalization_a}
A first possible generalization starts with the interpretation of eq. \ref{eq:prob_last_paper}, in which each gamma factor encodes a potential weight distribution. The number of weight distributions is fixed to the number of weights $N$. However, we can imagine to marginalize over the shape parameter of each gamma distribution with another discrete distribution, which models the Poisson uncertainty part of the CPD. If all weights are the same, this amounts to integrating out $N$. For practical reasons, we marginalize with a Poisson-gamma mixture distribution ($\mathrm{PG}$) instead of a Poisson distribution. We can write the result as
\begin{align}
&L_{\mathrm{Gen},(1)}(\theta) \\&= \int_{0}^{\infty} \frac{{e^{-\lambda}}{\lambda}^{k} }{k!} \cdot \left[ \left(\sum\limits_{i=0}^{\infty} \mathrm{G}_1(\lambda_{1};i+\alpha/N,1/w_{1}) \cdot \mathrm{PG}_1(i)\right) \ast \ldots \ast \left(\sum\limits_{j=0}^{\infty} \mathrm{G}_N(\lambda_{N};j+\alpha/N,1/w_{N}) \cdot \mathrm{PG}_N(j)\right)\right](\lambda) \ d\lambda  \\
&= \int_{0}^{\infty} \frac{{e^{-\lambda}}{\lambda}^{k} }{k!} \cdot \left[  \mathrm{GPG}_1(\lambda_{1})  \ast \ldots \ast  \mathrm{GPG}_N(\lambda_{N})\right](\lambda) \ d\lambda \label{eq:gen_a}
\end{align}
where $\mathrm{PG}(i)=\mathrm{PG}(i;1+\beta/N;1/v)$, i.e. the shape parameter of the Poisson-gamma mixture is $1+\beta/N$ and the rate parameter is $\frac{1}{v}$. We also define the mixture of the gamma distribution with the Poisson-gamma distribution as $\mathrm{GPG}=\sum\limits_{i=0}^{\infty} \mathrm{G}_N(\lambda_{N};i+\alpha/N,1/w_{N}) \cdot \mathrm{PG}_N(i)$. Here we have one mixture for each MC event. If two MC events have the same weight $w_1=w_2=w$, the convolution of two $\mathrm{GPG}$ distributions behaves similar as the convolution of two gamma distribution with the same rate parameter. We set $\beta=0$ and $v=1$ for simplicity and can write
\begin{align}
\sum\limits_{i=0}^{\infty} &\mathrm{G}_1(\lambda_{1};i+\alpha/N,1/w) \cdot \mathrm{PG}_1(i) \ast \sum\limits_{j=0}^{\infty} \mathrm{G}_2(\lambda_{2};j+\alpha/N,1/w) \cdot \mathrm{PG}_2(j) \\
&= \sum\limits_{i=0}^{\infty}  \sum\limits_{j=0}^{\infty} \mathrm{G}(\lambda_{1};i+j+2\cdot\alpha/N,1/w) \cdot \mathrm{PG}_1(i) \cdot \mathrm{PG}_2(j) \\
&= \sum\limits_{k=0}^{\infty}  \sum\limits_{j=0}^{k} \mathrm{G}(\lambda_{N};k+2\cdot\alpha/N,1/w) \cdot \mathrm{PG}_1(k-j) \cdot \mathrm{PG}_2(j) \\
&=\sum\limits_{k=0}^{\infty}  \mathrm{G}(\lambda_{N};k+2\cdot\alpha/N,1/w) \cdot \mathrm{PG}(k;2;1/v) 
\end{align}
When all weights are equal, we can therefore combine all $\mathrm{GPG}$-distributions in eq. \ref{eq:gen_a}, and obtain 
\begin{align}
\int_{0}^{\infty} \frac{{e^{-\lambda}}{\lambda}^{k} }{k!} \cdot \ \sum\limits_{i=0}^{\infty} \mathrm{G}(\lambda;i+\alpha,1/w) \cdot \mathrm{PG}(i;N;1/v) \ d\lambda
\end{align}
which represents a direct generalization of the formula involving the exact CPD (eq. \ref{eq:cpd_exact_int}) for equal weights discussed in section \ref{sec:exact_cpd}. Instead of the delta factor we have a gamma factor, and instead of the Poisson distribution we have a Poisson-gamma mixture distribution. The expression eq. \ref{eq:gen_a} therefore involves a certain generalization of a general compound distribution, where each of the continuous distributions $W_i$ is different.

The solution for $M$ different $\mathrm{GPG}$ distributions with different weights is derived in appendix \ref{appendix:gen_1} and yields

\begin{align}
&L_{\mathrm{Gen},(1)}(\theta) \\
&\stackrel[\substack{\alpha=0 \\ \beta=0 \\ v_j=1}]{}{=}\int_{0}^{\infty} \frac{{e^{-\lambda}}{\lambda}^{k} }{k!} \cdot \left[ \left(\sum\limits_{i=0}^{\infty} \mathrm{G}_1(\lambda_{1};i,1/w_{1}) \cdot \mathrm{PG}_1(i;1,1)\right) \ast \ldots \ast \left(\sum\limits_{j=0}^{\infty} \mathrm{G}_N(\lambda_{N};j,1/w_{N}) \cdot \mathrm{PG}_N(j;1,1)\right)\right](\lambda) \ d\lambda \\
&= \left(\prod\limits_{j=1}^{M} \cdot \left(\frac{1+w_j}{1+2w_j}\right)^{\langle k_{mc}\rangle_j}\right) \cdot \Delta_k
\end{align}
with
\begin{align}
\Delta_k=\frac{1}{k}\sum\limits_{i=1}^{k} \left[\left(\sum\limits_{j=1}^{M} \langle k_{mc,j} \rangle \cdot \left(\frac{2}{2+1/w_j}\right)^i - \langle k_{mc,j} \rangle  \cdot \left({\frac{1}{1+1/w_j}}\right)^i \right) \Delta_{k-i}\right]  \ \ \mathrm{and} \ \Delta_0=1
\end{align} for the case $\alpha=\beta=0$ and $v=1$ which we use for all later comparisons. The general solution with arbitrary parameters is given in appendix \ref{appendix:general_formulas}. Computationally, the calculation is only a little slower than eq. \ref{eq:poisson_gamma_mixture_simple_iterative} and still scales as $\mathcal{O}(N \cdot k^2)$. We tried the same calculation with a Poisson distribution as mixture instead of a Poisson-gamma distribution, but the result is computationally not as efficient. The Poisson-gamma distribution as the Posterior-Predictive distribution of observed counts is also well motivated, because it includes some uncertainty of the unknown underlying mean.

\subsection{Generalization (2) - Modeling the CPD for each dataset}
\label{sec:generalization_b}

In the second generalization we model the CPD of each dataset independently, and afterwards join all CPDs via convolution, instead of approximating all datasets with a combined CPD (see fig. \ref{fig:further_generalizations}). Before we do so, we first introduce a slightly different parametrization of the CPD compared to the one used by \cite{Argueelles2019}. Encoding mean and variance es defined in eq. (\ref{eq:z_mean}) and eq. (\ref{eq:z_variance}) to be mean and variance of a general gamma distribution with parameters $\alpha$ and $\beta$ we obtain
\begin{align}
\alpha&=\frac{(\mu_Z)^2 }{\mathrm{var}_Z} =\frac{(\mu_N\cdot \mathrm{E}[W])^2}{\mu_N \cdot (\mathrm{Var}[W] + (\mathrm{E}[W])^2)} \stackrel[\substack{sample \\ moments}]{}{=} \frac{(\frac{\mu_N}{k_{mc}} \sum_j w_j)^2}{\mu_N \cdot \left( \frac{1}{k_{mc}}\sum_j w_j^2\right)} 
= \frac{\mu_N \cdot  (\sum_j w_j)^2}{  k_{mc} \cdot \sum_j w_j^2}\\
\beta&
=\frac{\mu_N\cdot \mathrm{E}[W] }{\mu_N \cdot (\mathrm{Var}[W] + (\mathrm{E}[W])^2)} \stackrel[\substack{sample \\ moments}]{}{=}  \frac{\sum_j w_j}{\sum_j w_j^2}
\end{align} where we can observe that the unknown mean $\mu_N$ cancels out in $\beta$ but remains in $\alpha$. Choosing $\mu_N=k_{mc}+\frac{k_{mc}\cdot \sum_j w_j^2}{(\sum_j w_j)^2}$ results in $\alpha= \frac{(\sum_j w_j)^2}{\sum_j w_j^2}+1$ which is the choice that was made in \cite{Argueelles2019}. Instead of an additive nuisance parameter $a$ one can therefore choose a multiplicative parameter that is naturally given by $\mu_N$, which compared to the other choice has an intuitive interpretation as the unknown mean of the Poisson part of the CPD. We also do not have to use the sample based mean and variance for the weight distribution $W$, but can use other ways to encode the moments. For example a wider variance might be reasonable to model very few events (see section \ref{sec:incorporating_prior_info}) or we can use asymptotic Monte Carlo simulations to get the exact moments for crosschecks (see section \ref{sec:toy_mc}). The likelihood function with several such CPDs joined by convolution looks like  
\begin{align}
L_{\mathrm{Gen},2}=\int \mathrm{\mathbf{P}}(k;\lambda) \cdot \left[\mathrm{\mathbf{G}}(\lambda_1;\alpha_1, \beta_1)\ast \ldots \ast \mathrm{\mathbf{G}}(\lambda_{N_{src}};\alpha_{N_{src}}, \beta_{N_{src}}) \right](\lambda) d \lambda 
\end{align} where each gamma factor has its parameters $\alpha$ and $\beta$ encoded as described above.
Compared to the single CPD Ansatz by Arg\"uelles et al\cite{Argueelles2019}, the advantage is that large differences in statistics between datasets can be explicitly modeled in the respective $\mu_{N,j}$. It also has an interpretation as a direct probabilistic counterpart of the Barlow/Beeston \cite{Barlow1993} Frequentist Ansatz. In contrast to Barlow/Beeston, however, each CPD can take into account the full weight distribution, instead of averaging the weights. The analytic solution is known from \cite{Gluesenkamp2018} and given in appendix \ref{appendix:general_formulas}. It scales computationally as $\mathcal{O}(N_{\mathrm{src}} \cdot k^2+N)$. We can also introduce an effective version
\begin{align}
L_{\mathrm{Gen},2,\mathrm{eff}}=\int \mathrm{\mathbf{P}}(k;\lambda) \mathrm{\mathbf{G}}(\lambda;\alpha, \beta) d\lambda
\end{align} which is just a single CPD for all weights, similar to Arg\"uelles et al \cite{Argueelles2019}, but with the different parametrization of $\alpha$ and $\beta$ as introduced above. This allows to model per-dataset mean adjustments or weight distributions approximations as discussed later in section \ref{sec:incorporating_prior_info}.

\subsection{Generalization (3) - Marginalization of the unknown Poisson mean}
\label{sec:generalization_c}

Starting with the second generalization, we can go one step further and marginalize the unknown mean $\mu_{N,j}$ of the Poisson part of each CPD $j$ with another gamma distribution. The idea is to capture some of the uncertainty from ignorance about the unknown mean of the number of MC events per bin. We obtain
\begin{align}
&\begin{aligned}
 L_{\mathrm{Gen},(3)} &= \int_{0}^{\infty} \frac{{e^{-\lambda}}{\lambda}^{k} }{k!} \cdot \left[ \left(\int \mathrm{G}(\lambda_{1};\mu_1 \cdot \alpha_j^*,\beta_j)  \cdot \mathrm{G}(\mu_1; \frac{k_{mc,1}}{s},1/s) \ d \mu_1 \right) \ast \ldots \right. \\ &\left. \ast \left(\sum\limits_{j=0}^{\infty} \mathrm{G}(\lambda_{1};\mu_j \cdot \alpha_j^*,\beta_j) \cdot \mathrm{G}(\mu_j; \frac{k_{mc,N_{src}}}{s}, 1/s) \ d \mu_{ N_{src}}\right)\right](\lambda) \ d\lambda
\end{aligned}  \\
&= \int_{0}^{\infty} \frac{{e^{-\lambda}}{\lambda}^{k} }{k!} \cdot \left[ \mathrm{GG}_1 \ast \ldots \ast  \mathrm{GG}_{N_{src}} \right](\lambda) \ d\lambda \\
&= \mathrm{PGG}_1 \ast \ldots \ast \mathrm{PGG}_{N_{src}} \label{eq:convolutional_form_pgg}
\end{align} because we can write the mixture of a Poisson distribution with a convolution of gamma distributions as a discrete convolution of Poisson-gamma mixtures (see appendix \ref{appendix:general_formulas}. The scaling parameter $s$ has a natural choice of $s=1$. Choosing $s=2$ results in a gamma distribution with twice the variance but the same mean, and can be used to mimic the variance of a Poisson-gamma mixture distribution. The term $\alpha_j^*=\frac{ \mathrm{E}[W_j]^2}{\mathrm{Var}[W_j]^2}$ is again decoupled from the mean $\mu_{N_j}$, similar to the previuos generalization. To simplify terminology we use $\mathrm{GG}$ to denote a gamma distribution whose shape is marginalized by another gamma distribution, and $\mathrm{PGG}$ to denote a Poisson-gamma mixture distribution whose shape is marginalized with another gamma distribution. Several ways to calculate the solution exist, but we found the convolutional form (eq. \ref{eq:convolutional_form_pgg}) to be the most efficient way. Its computational complexity approximately scales as $\mathcal{O}(N_{src} \cdot k^2+N)$. The calculation of an individual $\mathrm{PGG}$ distribution is shown in appendix \ref{appendix:gen_3} and yields
\begin{align}
\mathrm{PGG_j}(k)=(\frac{1}{1+\beta_j})^k \cdot (1/s)^{k_{mc,j}/s}\sum\limits_{n=0}^{k} 
\frac{1}{ k!} \cdot {\alpha_j^*}^n  {k \brack n} 
\frac{\Gamma(k_{mc,j}/s+n)}{\Gamma(k_{mc,j}/s)(\gamma-\alpha_j^* \cdot \mathrm{ln}(\frac{\beta_j}{1+\beta_j}))^{k_{mc,j}/s+n}}
\end{align} with $\alpha_j^*=\frac{1}{k_{mc,j}}\frac{(\sum_j w_j)^2}{\sum_j w_j^2}$ and $\beta_j=\frac{\sum_j w_j}{\sum_j w_j^2}$ where we inserted sample mean and sample variance of the weight distributions. If not otherwise stated, we use $s=1$. The unsigned Stirling numbers of the first kind can be precomputed once via the recursion relation \cite{Abramowitz1970}
\begin{align}
{k \brack n} = (k-1){k-1 \brack n} + {k-1 \brack n-1}
\end{align} and then only require a table lookup.

\subsection{Overview}

An overview of the properties of all approaches is given in table \ref{table:summary_approaches}. The first column with weight uncertainty shows that most approaches approximate the underlying weight distribution to some extent. The exceptions are the approaches by Barlow/Beeston and by Bohm/Zech which approximate it essentially with a delta peak by averaging the weights. The second column shows which approaches include extra uncertainty for the unknown mean of the Poisson part of the CPD. Generalizations (1) and (3) are the only ones which do this, either by a discrete marginalization of the observed MC counts or by marginalizing the unknown mean with another probability distribution. The third column discusses the ability to adjust the expected mean for different datasets independently. The Frequentist approaches by Chirkin and by Barlow/Beeston in particular do this to some extent, as the profile likelihood fit implicitly varies parameters that can be interpreted as means of different Poisson components. However, in the second interpretation (see fig. \ref{fig:cpd_relationships}), this is not the case anymore. The fourth column indicates that only the probabilistic approaches converge to the absolute value of the Poisson likelihood as more Monte Carlo simulation is used. The Frequentist approaches only capture the shape, but their absolute values are meaningless.
The last column shows approximate computational complexities. While some approaches scale similarly, for example generalization (2) and (3), one has to remember that individually they require different mathematical operations, so in practice they might differ by a factor of a few. 

\begin{table}
	\setlength\extrarowheight{2.5pt}
	\centering
	\begin{tabular}{c|c|c|c|c|c}
		Name & \thead{different \\ weights} & \thead{extra Poisson \\ uncertainty} & \thead{Per-dataset \\ mean adjustment} & \thead{$N \rightarrow \infty$ \\ $\Rightarrow L\rightarrow L_{\mathrm{Poisson}}$ \\
			("probabilistic")}  & Computational complexity \\ \hline
		Bohm/Zech \cite{Bohm2012} & \xmark  & \xmark & \xmark & \xmark & $\mathcal{O}(N)$ \\
		Chirkin \cite{Chirkin2013b} & \cmark  & \xmark & \cmark (not manifest) & \xmark & $\mathcal{O}(N \cdot \mathrm{niter})$ ($N>1$)
		\\
		Barlow/Beeston \cite{Barlow1993} & \xmark & \xmark  & \cmark (not manifest) & \xmark & $\mathcal{O}(N + N_{src} \cdot \mathrm{niter})$ ($N_{\mathrm{src}}>1$) \\
		Arg\"uelles et al. \cite{Argueelles2019} & \cmark & \xmark  & \xmark & \cmark & $\mathcal{O}(N)$ \\
		Gl\"usenkamp \cite{Gluesenkamp2018} & \cmark & \xmark  & \cmark (not manifest) & \cmark & $\mathcal{O}(N \cdot k^2)$\\
		$L_{\mathrm{Gen},(1)}$ & \cmark & \cmark  & \cmark (not manifest) & \cmark & $\mathcal{O}(N \cdot k^2)$ \\
		$L_{\mathrm{Gen},(2)}$ & \cmark & \xmark  & \cmark  & \cmark & $\mathcal{O}(N_{\mathrm{src}} \cdot k^2+N)$ \\
		$L_{\mathrm{Gen},(2),\mathrm{eff}}$ & \cmark & \xmark  & \cmark (not manifest)  & \cmark & $\mathcal{O}(N)$ \\
		$L_{\mathrm{Gen},(3)}$ & \cmark & \cmark  & \cmark  &\cmark & $\mathcal{O}( N_{src} \cdot k^2+N)$
	\end{tabular}
	\caption{Summary table of different likelihood approaches that modify the Poisson likelihood. The number of Monte Carlo events  $N$, the number of datasets $N_{src}$ and the number of observed data $k$ are to be evaluated per bin. The Frequentist approaches (Barlow/Beeston and Chirkin) typically require multiple iterations ("niter") of a numerical solver when there are more than one source dataset (Barlow/Beeston) or events with different weights (Chirkin).}\label{table:summary_approaches} 
\end{table}

\section{Incorporating extra prior information}
\label{sec:incorporating_prior_info}

While the old and new formulas capture more uncertainty than the standard Poisson likelihood, the approximations are often biased, in particular if only a few MC events are present. This can be seen directly in section \ref{sec:generalization_b}: the mean $\mu_N$ is approximated by the observed MC counts and the weight distribution mean $\mu_W$ and variance $\mathrm{var}[W]$ approximated by the sample mean and variance of observed weights. If only a few events are present in a given bin, these estimates can be biased. For $\mu_N$ this happens when the Monte Carlo livetime is so low that several bins are empty, which indicates that the true mean $\mu_N$ is smaller than unity per bin. For $\mu_W$ and  $\mathrm{var}[W]$, this happens when the weight distribution is rather wide and only few events are present. Here we discuss a few approaches that can remedy these biases to some extent.

\subsection{Reducing bias on $\mu_n$}
\label{sec:mean_adjustment}
If the Monte Carlo livetime is very low it can happen that some bins are empty. Using the observed number of Monte Carlo events in the remaining bins that do see events can lead to an over estimation of the underlying mean. One strategy to circumvent this is to calculate the average number of Monte Carlo events over all bins that are eventually filled by the simulation. If this average is smaller than $1$, we subtract the difference from the sample mean of observed events in every bin. Once the average numer of Monte Carlo events per bin is above $1$, no correction is performed, so it only affects very low statistics. Importantly, it has to be done for every Monte Carlo dataset separately. If one has two datasets with very different statistics, and calculates this adjustment based on the combined average number of events per bin, it does not work. This is the reason generalizations (2) and (3) make this per-bin mean adjustment natural, because every dataset $j$ has its own mean $\mu_N,j$ that gets adjusted. However, it is in principle also possible to do this for most of the other approaches on a per event basis. This is what we call "not manifest" in table \ref{table:summary_approaches}.

\subsection{Reducing bias for $\mathrm{var}[W]$}
\label{sec:increased_weight_variance}

For very few events the sample variance might give a biased estimate of the true variance of the weight distribution. In particular, for a single event, the sample variance is zero, which is always an underestimation if the true weight distribution has a non-zero with. This happens whenever we have non-equal weights. Simply using the unbiased sample variance often does not help. One possibility to counteract this behavior is to assume that each weight represents a gamma distribution, and the total weight PDF is sum of individual gamma PDFs. The variance of a single gamma distribution with rate parameter $1/w$ is $w^2$, so the variance of a superposition of multiple gamma distributions is $\mathrm{var}[W]=\frac{1}{N}\sum_i w_i^2$. We will see in section \ref{sec:toy_mc} how this compares to the standard biased sample variance.

\subsection{Handling empty bins}

Empty bins must be handled with care. In particular, it is important to differentiate between datasets. For one dataset, a bin might be empty, while for another dataset it might be filled with many Monte Carlo events. If nothing is done about that, the resulting parameter estimation, which is likely connected only to one of the datasets, will be biased. In \cite{Barlow1993}, the strategy advocated was to add the dataset with the largest potential contribution based on all datasets that have no Monte Carlo events in the bin to add to the bin. However, this strategy adds discrete jumps to the likelihood, because for different parameters a different dataset might have higher contributions. Another strategy is discussed in \cite{Chirkin2013b}, which is just to add a constant "noise" contribution to all bins that are empty. In this strategy, however, no differentiation is made between datasets, which is bad for the reason stated above. The weight of the pseudo event is chosen to be the largest weight from all other bins, similar to the suggestion in \cite{Barlow1993}, but we do it for all datasets without any differentiation. The corresponding CPD approximations of these single "pseudo" Monte Carlo events again typically suffer the problem of an over estimation of the mean, which is discussed in the section \ref{sec:mean_adjustment}, and are almost by definition bias for the weight distribution. Therefore we typically combine the strategies for empty bins with the mean adjustment in section \ref{sec:mean_adjustment} and increased weight variance \ref{sec:increased_weight_variance}. Here we discuss two different strategies to incorporate empty bins. Both strategies basically fill up bins with "pseudo counts" for each dataset, which avoids the problem of jumps in the likelihood. The difference between the strategies involves the type of bins which are filled up.

\subsubsection{Strategy 1}

In the first strategy we only fill up bins that have at least one Monte Carlo event from any dataset. Once a bin is identified, we add  a single "pseudo" Monte Carlo event for every other dataset that has no events in it. 

\subsubsection{Strategy 2}

In the second strategy we fill up all bins that can potentially see Monte Carlo events, even those that see no events from any dataset yet. This is a more aggressive strategy, and leads to potentially larger bias for very low counts. The main aim of this strategy is to reduce likelihood fluctuations in the case when a new Monte Carlo simulation is run every parameter query (section \ref{sec:situation_2}). These fluctuations tend to be larger when totally empty bins are neglected, which happens by definition in strategy 1.

\section{Comparisons using Toy MC}
\label{sec:toy_mc}

\begin{figure}
	\centering
	\includegraphics{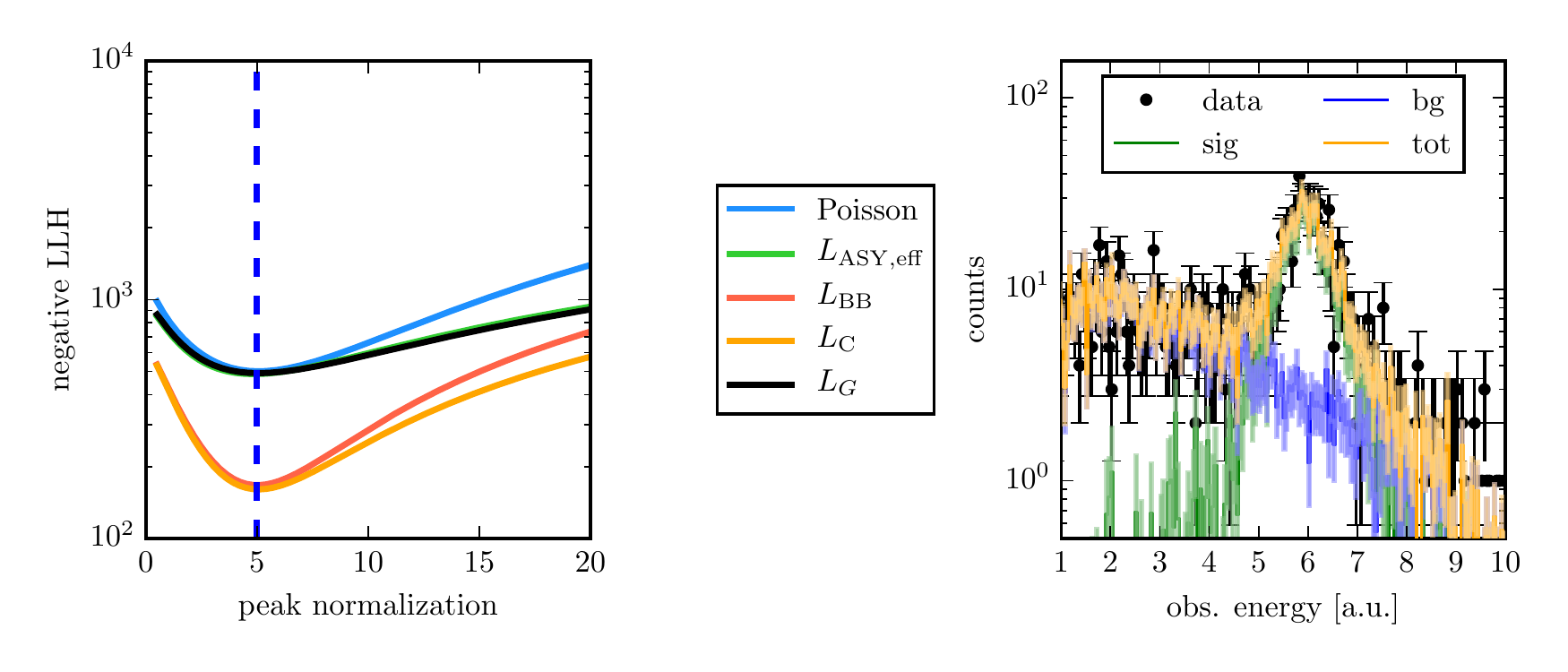}
	\caption{Example likelihood scans for various likelihood approaches from the literature in peak normalization (left). The non-standard likelihoods from other publications include $L_{\mathrm{ASY,eff}}$\cite{Argueelles2019},  $L_{\mathrm{BB}}$ \cite{Barlow1993}, $L_{\mathrm{C}}$ \cite{Chirkin2013b} and  $L_{\mathrm{G}}$\cite{Gluesenkamp2018}. The true parameter is indicated as the dashed verical line. One data realization and  Monte Carlo for the minimum of the Poisson scan is shown on the right.} \label{fig:test_explanation}
\end{figure}

To compare the various approaches, we will use a toy Monte Carlo setting of a falling background energy spectrum with a peaked signal spectrum on top. An artificial detector is simulated with energy smearing and energy-dependent detection efficiency and events are observed in some observable space. We might be interested in determining parameters related to the "signal" distribution like normalization and position, or parameters related to the "background" distribution, like normalization or spectral index. Figure \ref{fig:test_explanation} shows
 likelihood scans for the different standard approaches and the respective Monte Carlo and data distribution in observable space for the best-fit value of the Poisson likelihood. Assuming asymptotic behavior of the log-likelihood ratio $\lambda=-2 \cdot \Delta \mathrm{LLH}(\hat{p}, p_{0}) \sim \chi_\nu^2$ \cite{PDG2018}, one can determine the significance of exclusion of a given parameter value $p_0$, given $\hat{p}$ the best fit value. In the following tests, instead of using the log-likelihood ratio in the usual way, we show the bias of the log-likelihood ratio $\lambda$ (in $\sigma$ equivalents) for a specific data realization compared to infinite Monte Carlo statistics for a given true parameter $p_0$. We also show the expected coverage vs. the actual coverage by evaluating 500 Toy experiments for a given simulated livetime. 
\subsection{Situation 1: Likelihood ratio bias and coverage}

\subsubsection{Equal weight test}

\begin{figure}
	\centering
	{\includegraphics{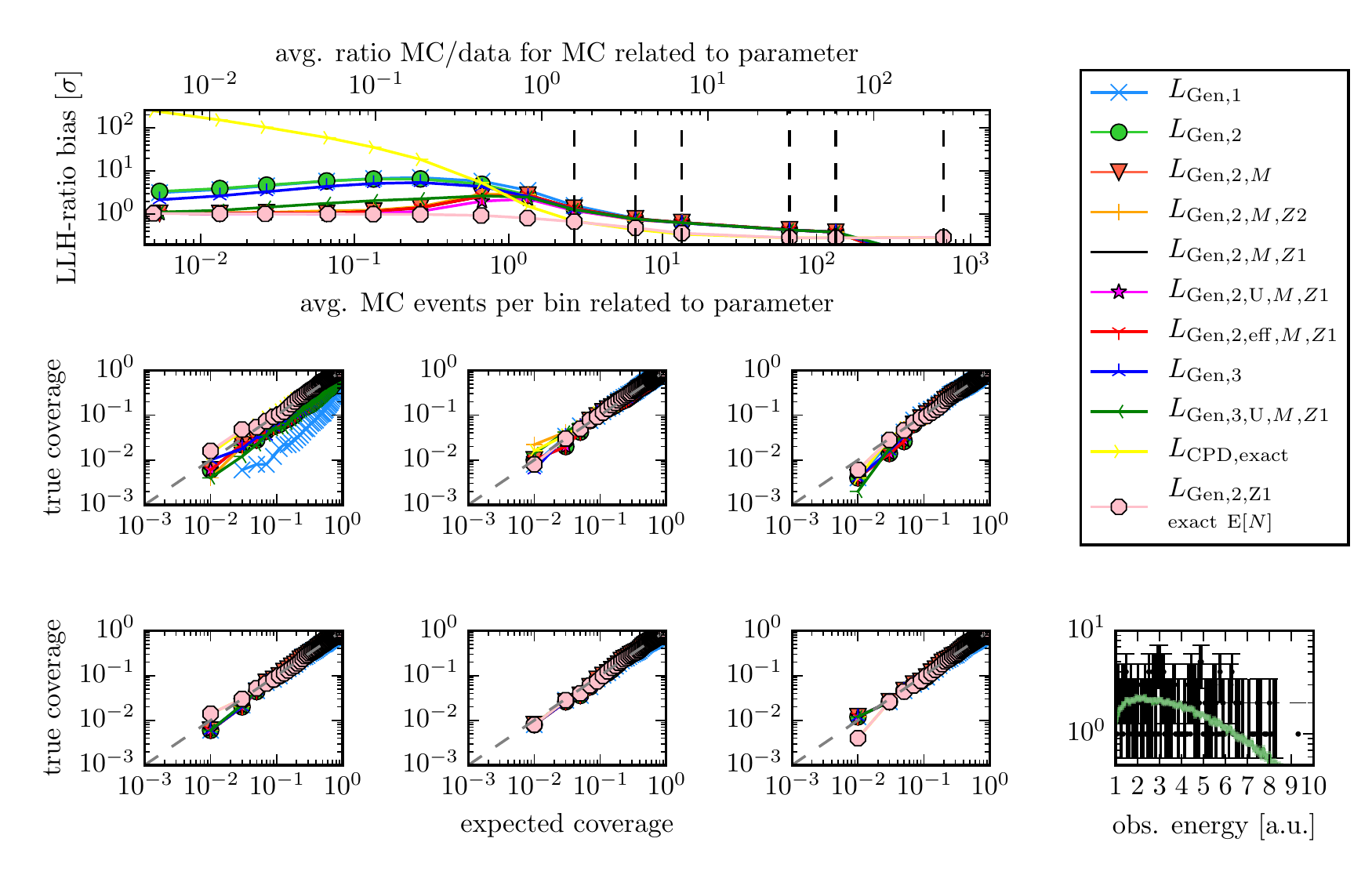}}
	\caption{Bias of the log-likelihood ratio $\lambda$ and coverage in dependence of the background dataset statistics. The signal is not present. The parameter of interest is the normalization of the background dataset. $Z1$ denotes strategy 1 for "zero MC", $Z2$ strategy 2 for "zero MC", $U$ denotes more "unbiased" weight variance estimation and $M$ denotes shifted mean based on average MC events in all bins. The coverage is shown at 6 statistic levels (in order) which are indicated as dashed vertical lines in the upper plot. In the lower right plot a particular data realization and a MC realization at the highest statistic level is shown for the observable space.} \label{fig:e1}
\end{figure}

\begin{figure}
	\centering
	{\includegraphics{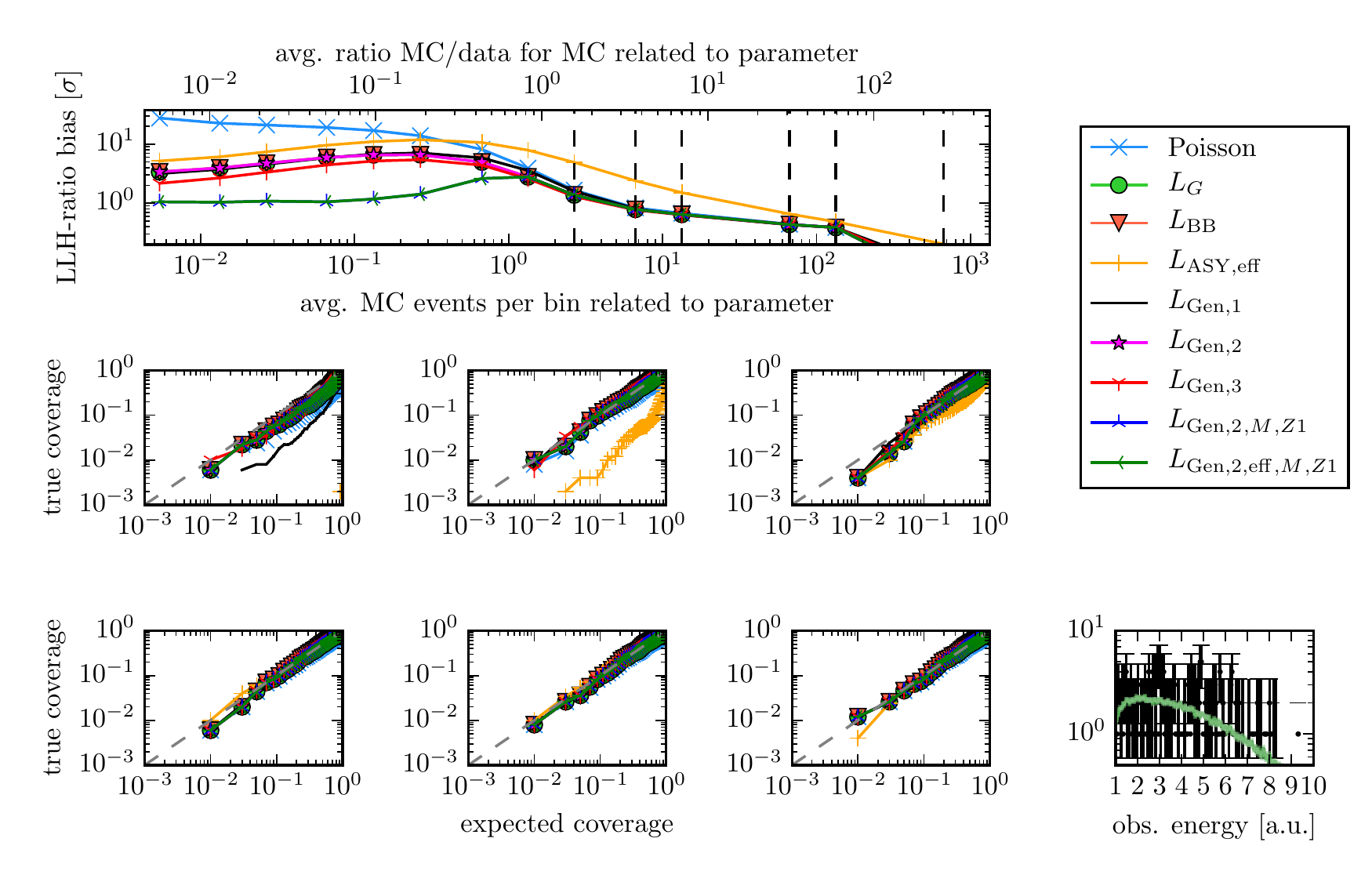}}
	\caption{Bias of the log-likelihood ratio $\lambda$ and coverage in dependence of the background dataset statistics. The signal is not present. The parameter of interest is the normalization of the background dataset. $Z1$ denotes strategy 1 for "zero MC", $Z2$ strategy 2 for "zero MC", $U$ denotes more "unbiased" weight variance estimation and $M$ denotes shifted mean based on average MC events in all bins. The coverage is shown at 6 statistic levels (in order) which are indicated as dashed vertical lines in the upper plot. In the lower right plot a particular data realization and a MC realization at the highest statistic level is shown for the observable space. The non-standard likelihoods from other publications include $L_{\mathrm{ASY,eff}}$\cite{Argueelles2019},  $L_{\mathrm{BB}}$ \cite{Barlow1993}, $L_{\mathrm{C}}$ \cite{Chirkin2013b} and  $L_{\mathrm{G}}$\cite{Gluesenkamp2018}.} \label{fig:e2}
\end{figure}

The simplest test one can perform is a fit involving only the background dataset. We use a simple weighting function that just scales normalization (see fig. \ref{fig:mc_illustration}) and all weights for the background dataset are the same. In this particular case of equal weights, we can calculate the CPD exactly (see section \ref{sec:exact_cpd}). We perform a large number of Monte Carlo simulations at each statistical level to obtain a good estimator of the true mean $\mu_N$, while $\mu_W=w$ and $\mathrm{var}[W]=0$ are known since we deal with equal weights. We compare this "exact" CPD likelihood with the new generalizations (1), (2), and (3), and a particular form of generalization (2) where we replace the unknown mean $\mu_N$ not by the observed number of MC events $k_{mc}$, but also by the true mean ("exact $\mathrm{E}[N]$") obtain from a large number of MC realizations. This can be understood as a continuous approximation of the exact CPD. The comparison of these different likelihoods in terms of $\lambda$-bias and coverage is shown in fig. \ref{fig:e1}.  
The first surprising observation is that the exact CPD ($L_{\mathrm{CPD,exact}}$) has a large bias at low average MC events per bin, although one would expect that the exact CPD works at any level of Monte Carlo statistics. This probably has to do with the multi-modal structure of the exact CPD. The continuous approximation ($L_{\mathrm{Gen,2,Z1, exact E[N]}}$) on the other hand is basically unbiased everywhere. We find that generalization (1), which is similar in nature to the exact CPD, performs not so good, which is related to the fact that the exact CPD itself performs badly, and we only plot it for the simple case without any modifications. Generalization (2) and generalization (3), which are again continuous approximations, perform better and are almost equivalent in this scenario. We can also observe how the mean adjustment is important to reduce bias, while the strategy for empty bins or using a more generous estimate for the weight variances do not change things too much in this particular scenario.  Comparing the approach using the adjusted mean ($L_{\mathrm{Gen,2,M}}$) with the exact formula ($L_{\mathrm{Gen,2,Z1, exact E[N]}}$), we see that in the intermediate statistics regime there still is some bias in the likelihood ratio that is not removed by the mean adjustment strategy applied here. One could think about addressing this problem by some modification of the strategy, but we do not further test it here.  Figure \ref{fig:e2} shows the best performing cases of generalization (2) and generalization(3) in comparison to older likelihood approaches. In general, the new approaches show lower bias and better coverage over a wide range of MC statistics. 

\subsubsection{Background dataset with limited statistics}
\label{sec:toymc_different_dataset_statistics}
\begin{figure}
	\centering
	{\includegraphics{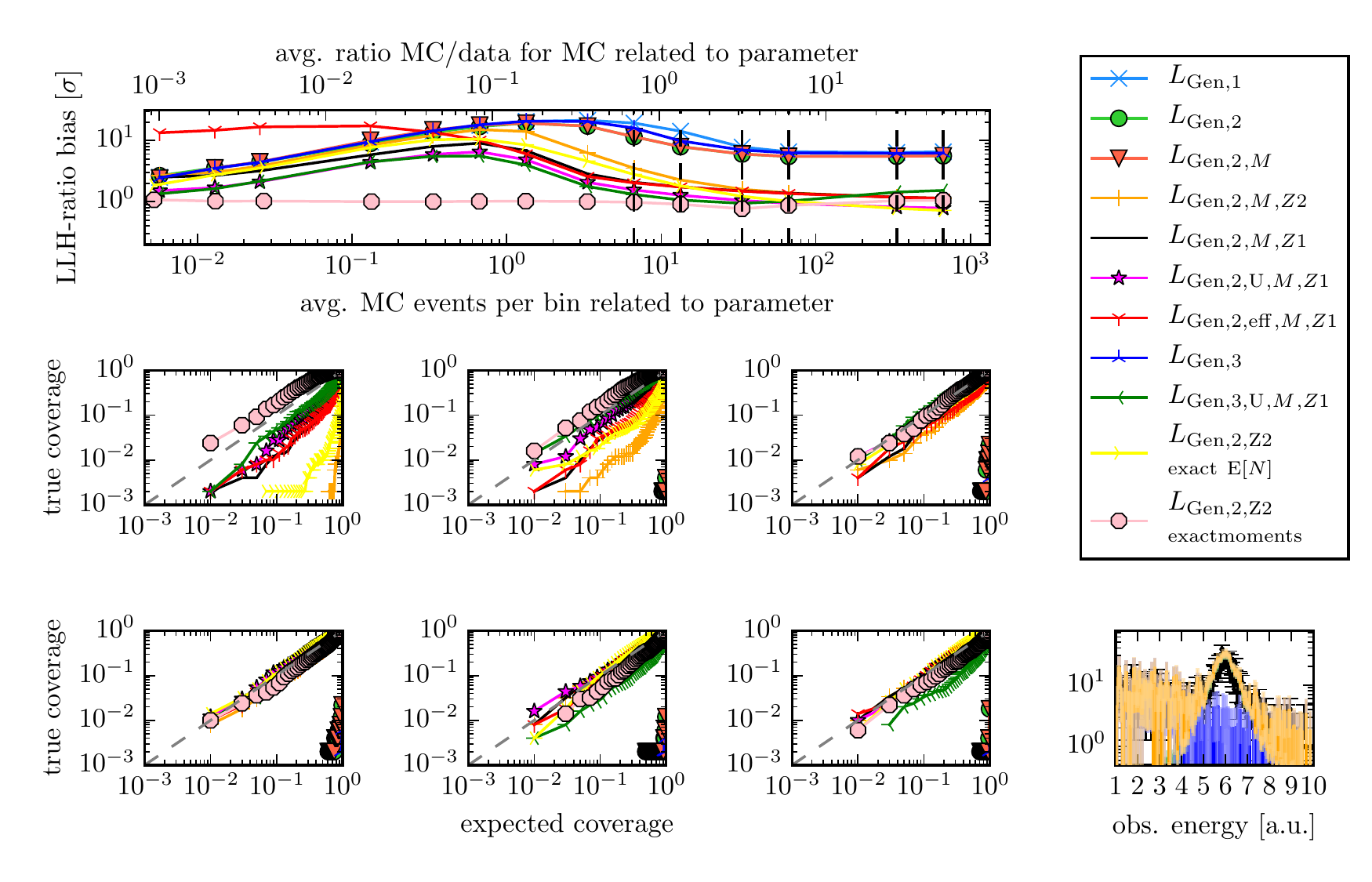}}
	\caption{Bias of the log-likelihood ratio $\lambda$ and coverage in dependence of the signal dataset statistics. The background statistics stays the same. The parameter of interest is the normalization of the peak in the signal dataset. $Z1$ denotes strategy 1 for "zero MC", $Z2$ strategy 2 for "zero MC", $U$ denotes more "unbiased" weight variance estimation and $M$ denotes shifted mean based on average MC events in all bins. The coverage is shown at 6 statistic levels (in order) which are indicated as dashed vertical lines in the upper plot. In the lower right plot a particular data realization and a MC realization at the highest statistic level is shown for the observable space.} \label{fig:mid_1}
\end{figure}

In this test setting we fix the background dataset at limited statistics, while the signal statistics are increased step-by-step. We fit the normalization of the signal peak. Such a situation with limited background statistics can happen if hard cuts are applied in the data selection process and computational resources are not sufficient to still have enough simulated background events surviving at final level. We compare various combinations of treating empty bins with incorporating prior information. Again we first compare various strategies with the new formulas as shown in fig. \ref{fig:mid_1}. 
 We can obtain the exact moments from high statistics simulations and use these moments at the respective level, which results in a unbiased likelihood ratio at all statistic levels for crosschecks (($L_{\mathrm{Gen,2,Z1, exact moments}}$)). If we use the exact Poisson mean, but use sample estimates for the weight moments ($L_{\mathrm{Gen,2,Z1, exact E[N]}}$), a bias becomes visible. We can compare this result to the likelihood of generalization (2) with mean adjustment, which shows the mean adjustment is nearly as good as using the exact Poisson mean. We can also see that filling up the empty bins with pseudo counts is necessary here. All likelihood approaches that do not fill up empty bins (no "ZB") end up heavily biased even with infinite signal Monte Carlo, only because the background Monte Carlo stays limited. Additionally, usage of a more flexible weight variances estimator ("U") can be helpful. For generalization (3), though, having a more unbiased estimator ("U") seems to introduce a slight bias at high statistics, which we have not fully understood yet. Potentially the total variance of the estimators is too large here (compare also fig. \ref{fig:mid_2_bias} in appendix \ref{appendix:suppl_plots} without "U"). Lastly, the effective version of generalization (2) does not seem to perform well in this scenario for very low statistics, but for higher statistics it approaches the standard formula of generalization (2). In figure \ref{fig:mid_2_bias}, which we show in appendix \ref{appendix:suppl_plots}, we compare again generalization (2) against older approaches from the literature and also show the maximum likelihood estimator (MLE) bias and variance. We see again that all approaches from the literature fail when empty bins are neglected. Interestingly, the standard Poisson likelihood performs quite good in a low-statistic certain region, which has to do with very low MLE bias compared to all modifications. 

\subsubsection{Other scenarios}

We also tested several other scenarios, whose results we show in appendix \ref{appendix:suppl_plots}. One scenario deals with more statistics and a smaller relative peak where we fit the peak normalization. The results are shown in fig \ref{fig:highbg_lowpeak_bias}. This situation is more akin to the scenario that was studied by the authors in \cite{Argueelles2019} ("$L_{\mathrm{ASY,eff}}$"). We see that our likelihoods with mean adjustment perform better, in particular the effective version of generalization (2) ("$L_{\mathrm{Gen2,eff}}$") , which shows that a different parametrization is really crucial. In yet another scenario we increase the livetime of background and signal simultaneously and fit the spectral index (see fig. \ref{fig:sifit_bias}). Again, the results are similar.

\subsection{A new MC set every minimization step}
\label{sec:situation_2}
\begin{figure}
	\centering
	{\includegraphics{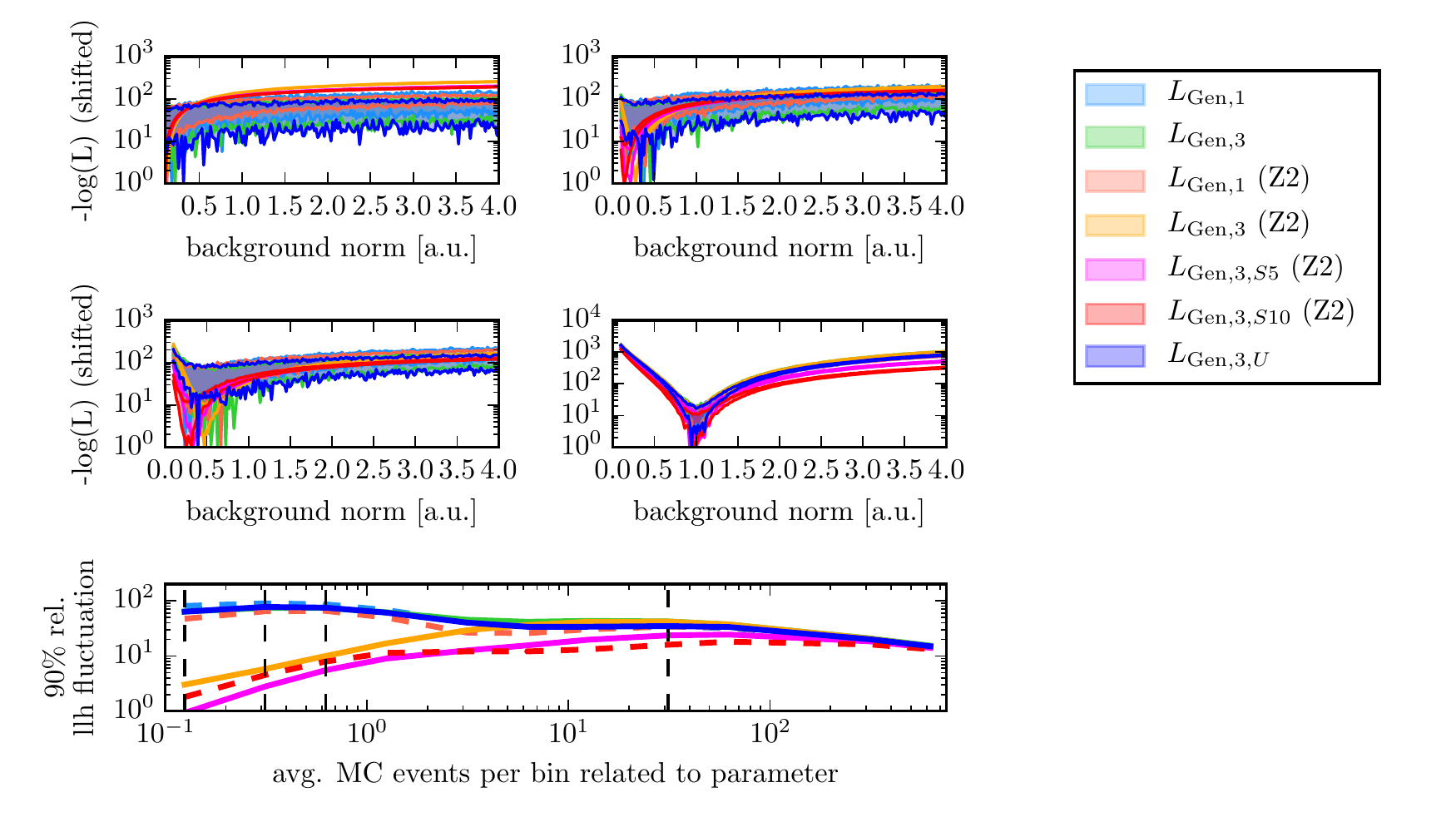}}
	\caption{Stochastic scans of the likelihood functions at four different statistic levels (top). Shown is the central $90 \%$ interval of 50 realizations, shifted in absolute value for visualization so one can directly compare their shapes. In the bottom part we show the average of such likelihood fluctuations in dependence of the simulated Monte Carlo statistics. Vertical lines indicate which statistical levels are shown in the four plots above. $Z2$ denotes strategy 2 for "zero MC", $U$ denotes more "unbiased" weight variance. Different parameters values of $S$ correspond to slightly different behavior of generalization (3).} \label{fig:fluctuations_1}
\end{figure}

\begin{figure}
 	\centering
 	{\includegraphics{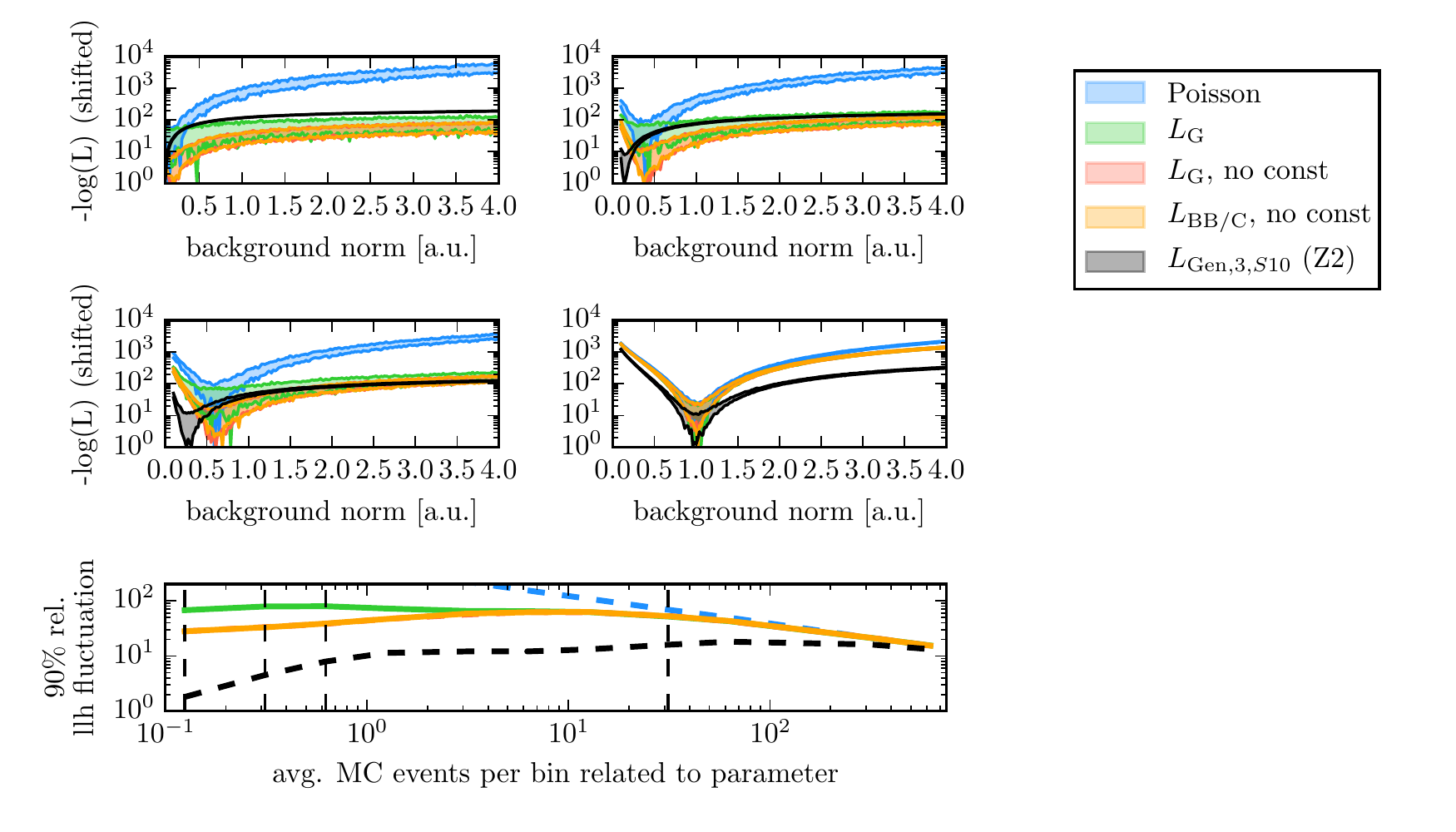}}
 	\caption{Stochastic scans of the likelihood functions at four different statistic levels (top). Shown is the central $90 \%$ interval of 50 realizations, shifted in absolute value for visualization so one can directly compare their shapes. In the bottom part we show the average of such likelihood fluctuations in dependence of the simulated Monte Carlo statistics. Vertical lines indicate which statistical levels are shown in the four plots above. $Z2$ denotes strategy 2 for "zero MC". Different parameters of s relate to parameterizations of generalization (3). The older formulas are the approaches from  \cite{Barlow1993} \cite{Chirkin2013b} ("$L_{\mathrm{BB/C}}$"), which are similar for equal weights, and from \cite{Gluesenkamp2018} ("$L_{\mathrm{G}}$").} \label{fig:fluctuations_2}
 \end{figure}

In certain applications we might simulate a new set of MC events for every successive parameter call. This is a usual technique in IceCube to determine the ice properties or to perform expensive particle reconstructions \cite{Chirkin2013}. Even when the parameters are the same, a new set of MC events will have different weights and a different simulated count per bin (see \ref{sec:introduction} fig. \ref{fig:mc_illustration}). In this situation, the uncertainty is directly visible in fluctuations of the corresponding likelihood function. In fig. \ref{fig:fluctuations_1} we compare various versions of generalization (1) and generalization (3), which directly model this uncertainty within the PDF, for a likelihood scan of the normalization of the background dataset.
In this scenario all weights are the same.
We test different parameter values of the scaling parameter $S$ from generalization (3) (see section \ref{sec:generalization_c}). A larger scaling parameter leads to a larger uncertainty being considered in the likelihood function, and correspondingly the fluctuations go down. However, we can observe that the likelihood bias goes up. A useful strategy might therefore be to start with a large parameter value of $S$ and decrease it during the optimization process. Similarly, we could use generalization (1) and modify the parameter $v$ (see section \ref{sec:generalization_a}) and introduce another scaling factor for the respective shape parameters of the Poisson-gamma mixtures. This should have a similar effective behavior. Additionally, we see that filling up the empty bins (Z2) is a second ingredient which is important. In fig. \ref{fig:fluctuations_2} we compare the best performing generalization (3) with previous approaches. Here we observe that it is also possible to lower relative fluctuations by removing constant parts from the likelihood function which lowers the absolute values, and thereby also the relative fluctuations. The approaches described in \cite{Chirkin2013} \cite{Barlow1993} \cite{Gluesenkamp2018} yield exactly similar fluctuations in this case. However, they can not compete with a direct uncertainty modeling. 
\section{Conclusion}

The limited statistics of Monte Carlo simulation leads to extra uncertainty that has to be taken into account for correct parameter estimation. In this paper, we discussed several solutions from the literature that modify the Poisson likelihood to tackle this problem. We showed that all the approaches of this type can be interpreted as approximating the underlying probability distribution of the sum of weights (the CPD) and then use this approximation to solve a marginalization integral over the Poisson mean. In probabilistic approaches, the integral is calculated exactly. In Frequentist approaches, the integrand is optimized over the mean, not marginalized. This is an alternative viewpoint to the prevailing interpretation of Frequentist approaches (\cite{Barlow1993} \cite{Chirkin2013b}), which is that these simply treat the Monte Carlo data similar to real data in the likelihood function. 

In the new perspective, the central question is really how to approximate the mean $\mu_N$ of the Poisson part of the CPD, as well as the mean $\mu_W$ and variance $\mathrm{var}[W]$ of the continuous "weight part" of the CPD. 
We introduced three new generalizations of existing probabilistic approaches that are well-motivated in this context. In generalization (1), we marginalize each gamma distribution from the probabilistic Ansatz in \cite{Gluesenkamp2018} with another Poisson-gamma mixture distribution. For equal weights, this distribution involves a marginalization of a Poisson-gamma mixture compound distribution, which can be viewed as a generalization of the exact CPD with more variance. In generalization (2), we model the CPD of each dataset independently with a gamma distribution, which are then convolved to form the total CPD. It can be viewed as a generalization of the approach by \cite{Argueelles2019} in combination with the convolutional formula introduced in \cite{Gluesenkamp2018}. It is also the direct probabilistic counterpart to the approach by Barlow et al \cite{Barlow1993}. Each gamma distribution is parametrized by the relevant CPD moments $\mu_N$, $\mu_W$, $\mathrm{var}[W]$ of a given dataset, which are usually taken to be the sample moments, but they don't have to be. We also discuss an effective version of generalization (2) which encodes a single CPD for all datasets, similar to \cite{Argueelles2019}, but with a different parametrization that allows to incorporate extra Prior information in a more natural way. In generalization (3), we marginalize the unknown mean $\mu_N$ of each CPD from generalization (2) with yet another gamma distribution to incorporate extra sampling uncertainty directly into the PDF. All of the new probability distributions are analytically solvable. 

In the last part of the paper we test these new formulas in a Toy MC parameter estimation setting with two datasets, a background dataset and signal dataset. We find that generalization (1) performs quite poor in these tests in terms of coverage and likelihood ratio bias. We also surprisingly find that the exact CPD performs worse than a continuous approximation with a gamma distribution, especially at very low statistics. We suspect that this has to do with the multi-modal form of the exact CPD. This explains why generalization (1) performs so poor for correct coverage, because it is just a slightly more general version of the exact CPD in the equal weights setting and has similar multimodal structure. Generalizations (2) and (3), which are based on continuous approximations of the overall CPD, perform better in all coverage tests.

If the moments are known exactly, a continuous approximation of the CPD as it is done in generalization (2) is in a certain sense optimal: it never undercovers  at *any* level of Monte Carlo livetime. Achieving good coverage therefore reduces to finding good approximations to the moments $\mu_N$, $\mu_W$ and $\mathrm{var}[W]$ - the better these approximations, the better the coverage properties. Here, the usual sample moments $\mu_N=k_{mc}$, $\mu_W=\frac{1}{k_{mc}}\sum_i w_i$, $\mathrm{var}[W]=\frac{1}{k_{mc}}\sum_i w_i^2$ are often biased which leads to undercoverage. 

We can remedy this problem to some extent by incorporating extra Prior information about the Monte Carlo into these moments. One of these is a "mean adjustment" based on the average Monte Carlo count in all bins. The mean, and thereby also the mean adjustment, can be interpreted as a hyperparameter similar to the parameter $a$ in the effective likelihood discussed in \cite{Argueelles2019}, and we show how they are precisely related. In contrast to the choice of an effective likelihood with $a=1$ discussed in \cite{Argueelles2019}, the mean adjustment is conceptually easier to interpret and gives better results in all test scenarios. 

We also discuss a certain strategy how to treat bins with zero Monte Carlo events for a given dataset. In contrast to previous suggestions from the literature (see \cite{Barlow1993}) how to treat such scenarios, we find that a good choice is to fill up bins with pseudo counts such that every dataset has at least a single event in every bin, indifferent to the relative weights between datasets. This has the advantage that no jumps in the likelihood can appear. 

In a final test we studied the scenario of new simulation Monte Carlo simulation every parameter step, as it happens for example in the application of direct Cerenkov photon simulation in the Icecube observatory \cite{Chirkin2013}. Here, we observed that generalization (1) and generalization (3), which take into account the Poisson uncertainty from repeated sampling steps, greatly reduce fluctuations in successive likelihood evaluations compared to all other approaches. Given that generalization (3) has better properties in terms of log-likelihood ratio bias, it seems to be better suited than generalization (1) in this use case.

Our overall recommendation for parameter estimation situations can be summarized as follows:
\begin{enumerate}
	\item Normal use case of weighted simulation: Use generalization (2) with strategy 1 for empty bins and mean adjustment. If the computational cost is too high for the application, use the effective version of generalization (2).
	\item Re-simulation every parameter query: Use generalization (3) with strategy 2 for empty bins and a scaling parameter $s>1$ to minimize fluctuations. Potentially decrease the scaling paramter during the minimization process to unity to decrease bias.
\end{enumerate}

 Besides the tests with exact moments using perfect knowledge from "infinite Monte Carlo",  it was never possible to reduce the bias of the log-likelihood ratio completely at all levels of simulated livetime. Blindly trusting a formula can therefore lead to trust in a systematically wrong result if log-likelihood ratio values are directly translated into significances. The way out are ensemble-based test statistic constructions. These might again be problematic because Monte Carlo is expensive to produce. If ensemble-based constructions are possible, however, we recommend to always use the new formulas because they capture the underlying statistics of the CPD. In general, it is always a good idea to calculate the log-likelihood ratio with multiple likelihood functions, at least also the standard Poisson likelihood, and conclude from the differences between the different approaches if enough Monte Carlo statistics have been produced. Probabilistic approaches allow to additionally compare absolute likelihood values. If those absolute values disagree substantially, for example between the Poisson likelihood and the modified likelihood in use, it is a strong hint that one must be careful on the interpretation of the result. Implementations of the new formulas can be found on \href{http://www.github.com/thoglu/mc\_uncertainty}{\texttt{http://www.github.com/thoglu/mc\_uncertainty}}.

\section*{Acknowledgements}
We would like to thank Carlos Arg\"uelles, Marty Cohen, Austin Schneider and Tianlu Yuan  for stimulating discussions.

\appendix

\section{Derivation generalization (1)}
\label{appendix:gen_1}
This section computes the generalization described in section \ref{sec:generalization_a}. 

\subsection{Equal weights}

We start with equal MC weights for simulated events in a given bin. In this case, the generalized per-bin likelihood function is given by suitable marginalization as \cite{Gluesenkamp2018} 
\begin{align}
L_{\mathrm{bin, eq.}} &=
\mathrm{E}\left[\frac{{e^{-\lambda}}{\lambda}^{k} }{k!}\right]_{{\mathrm{G}(\lambda; k_{mc}+\alpha, 1/w)}} \nonumber \\
& =\frac{ \Gamma(k+k_{mc} +\alpha)}{\Gamma(k_{mc} + \alpha) \cdot k!} \cdot \frac{ (1/w)^{k_{mc}+\alpha}}{  (1+1/w)^{k+k_{mc}+\alpha}} \label{eq:poisson_mixture_equal_classical} \\
&= \mathrm{PG}(k;k_{mc}+\alpha, 1/w) \label{eq:simple_gp_mixture}
\end{align}
where the result is a standard gamma-Poisson mixture $\mathrm{PG}(k; k_{mc}+\alpha,1/w)$ distribution. For more information on the parameter $\alpha$, which can be seen as a hyperparameter parametrizing a gamma prior, in particular its special value $\alpha=0$, see \cite{Gluesenkamp2018}. Let us now extend this construction by further marginalizing out the discrete counts $k_{mc}$ via
\begin{align}
L_{\mathrm{bin, (1), eq.}} &= \mathrm{E}\left[L_{\mathrm{bin, eq.}}\right]_{{\mathrm{PG}(k_{mc}; \langle k_{mc}\rangle + \beta, 1/v)}} \nonumber \\ 
&= \sum\limits_{k_{mc}=0}^{\infty}
\mathrm{PG}(k; k_{mc}+\alpha, 1/w) \cdot \mathrm{PG}(k_{mc}; \langle k_{mc}\rangle + \beta, 1/v) \label{eq:discrete_marginal_equal} \\
&\equiv \mathrm{PGPG}(k;\alpha;1/w;\langle k_{mc}\rangle + \beta, 1/v)
\end{align}
and obtain a new likelihood function $L_{\mathrm{\textbf{P}, finite, twice,eq.}}$
that has been marginalized twice, whose probability distribution we abbreviate as $\mathrm{PGPG}$, because we marginalize a Poisson-gamma mixture with another Poisson-gamma mixture. This second marginalization is discrete, and happens with another gamma-Poisson mixture $\mathrm{PG}(k_{mc}; \langle k_{mc}\rangle+\beta,1/v)$ which is similar in structure to the Poisson mixture in eq. \ref{eq:poisson_mixture_equal_classical}, but with different parameters \footnote{We could have used a standard Poisson distribution for marginalization instead of a Poisson-gamma mixture. However, it turns out that such an Ansatz gives a more complicated result for general weights which are discussed in the next section. In particular, essential singularities appear which make the end result harder to compute.}. We will later set the "count weight" $v$ to $v=1$, but for now keep it in this general form. A weight of unity makes sense since we are now interested in marginalizing pure counts, which are unweighted by construction. Another nuisance parameter $\beta$ is introduced to handle Prior freedom, similar to the introduction of $\alpha$ in \cite{Gluesenkamp2018}.
In eq. \ref{eq:simple_gp_mixture} the variable  $k_{mc}$ denotes the observed number of MC events, while in the marginalized expression (eq. \ref{eq:discrete_marginal_equal}) the variable that is integrated out is $k_{mc}$ and the observed MC counts as $\langle k_{mc} \rangle$. Equation \ref{eq:discrete_marginal_equal} can be solved using Egorychev rules (\cite{Egorychev1984} p. 19) via
\begin{align}
L_{\mathrm{bin}, (1), eq.}&=\sum\limits_{k_{mc}=0}^{\infty}
\mathrm{PG}(k; k_{mc}+\alpha, 1/w) \cdot \mathrm{PG}(k_{mc}; \langle k_{mc}\rangle+\beta, 1/v) \\
&= \sum\limits_{k_{mc}=0}^{\infty} \frac{ \Gamma(k+k_{mc}+\alpha)}{\Gamma(k_{mc}+\alpha) \cdot k!} \cdot \frac{ (1/w)^{k_{mc}+\alpha}}{  (1+1/w)^{k+k_{mc}+\alpha}} \frac{\Gamma(k_{mc}+\langle k_{mc}\rangle+\beta)}{\Gamma(\langle k_{mc}\rangle+\beta) \cdot k_{mc}!} \cdot
\frac{ (1/v)^{\langle k_{mc}\rangle+\beta}}{  (1+1/v)^{k_{mc}+\langle k_{mc}\rangle+\beta}}
\\
&=
\begin{aligned} \underbrace{\left(\frac{1}{1+1/w}\right)^{k} \left(\frac{1}{1+w}\right)^{\alpha} \cdot \left(\frac{1}{1+v}\right)^{\langle k_{mc}\rangle+\beta}}_{\equiv K} \\ \cdot \sum\limits_{k_{mc}=0}^{\infty} \Big(\underbrace{\frac{v}{(1+v)(1+w)}}_{\equiv C}\Big)^{k_{mc}}\frac{1}{2\pi i} \oint\limits_{|z_1|=\epsilon} &\frac{(1+z_1)^{k+k_{mc}+\alpha-1} }{ z_1^{k+1}} dz_1 \frac{1}{2\pi i} \oint\limits_{|z_2|=\epsilon}\frac{ (1+z_2)^{k_{mc}+\langle k_{mc}\rangle+\beta-1}} { z_2^{k_{mc}+1}} dz_2 
\end{aligned} \label{eq:contant_def_1} \\
&=
\begin{aligned} K \cdot \frac{1}{2\pi i}\oint\limits_{|z_1|=\epsilon} \frac{(1+z_1)^{k+\alpha-1} }{ z_1^{k+1}} \\
\cdot \sum\limits_{k_{mc}=0}^{\infty} C^{k_{mc}} \cdot \frac{1}{2\pi i} \oint\limits_{|z_2|=\epsilon}& (1+z_2)^{\langle k_{mc}\rangle+\beta-1} \cdot \Big[(1+z_1)(1+z_2)\Big]^{k_{mc}} \cdot z_2^{-k_{mc}-1} dz_2 dz_1 \end{aligned}  \\
&= K \cdot \frac{1}{2\pi i}\oint\limits_{|z_1|=\epsilon} \overbrace{\frac{(1+z_1)^{k+\alpha-1} }{ z_1^{k+1}}}^{\mathrm{numerator} \equiv f(z_1)} \cdot \underbrace{\Big(\frac{1}{1-C(1+z_1)}\Big)^{\langle k_{mc}\rangle+\beta}}_{\equiv g(z_1)} dz_1 =  K \cdot \frac{1}{2\pi i}\oint\limits_{|z_1|=\epsilon} \frac{f(z_1)}{ z_1^{k+1}} \cdot g(z_1) dz_1  \label{eq:pgpg_contour} 
\end{align}
Using the Taylor expansion of $f(z_1)$ and $g(z_1)$ around $z_1=0$, we can combinatorially combine the relevant terms to contribute to the residue of the contour integral at $z_1=0$. The result is then readily written down as
\begin{align}
L_{\mathrm{bin}, (1), eq.}&=K \cdot \sum\limits_{n=0}^{k} \frac{1}{n!}f^{(n)}(0)\cdot \frac{1}{(k-n)!}g^{(k-n)}(0)\\
&= K \cdot \Big(\frac{1}{1-C}\Big)^{\langle k_{mc}\rangle+\beta}\cdot \Big(\frac{C}{1-C}\Big)^k \sum\limits_{n=0}^{k} \frac{\Gamma(k+\alpha)}{n! \Gamma(k-n+\alpha)} \frac{\Gamma(\langle k_{mc}\rangle + \beta+k-n)}{(k-n)! \Gamma(\langle k_{mc}\rangle + \beta)}  \cdot \Big(\frac{1-C}{C}\Big)^{n} 
\end{align}
where the constants $K$ and $C$ are defined in eq. \ref{eq:contant_def_1} as $K=\left(\frac{1}{1+1/w}\right)^{k} \left(\frac{1}{1+w}\right)^{\alpha} \cdot \left(\frac{1}{1+v}\right)^{\langle k_{mc}\rangle+\beta}$ and $C=\frac{v}{(1+v)(1+w)}$.

\subsection{General weights}
\label{appendix:gena_general_weights}

For general weights we start with eq. \ref{eq:gen_a} and rewrite it as 
\begin{align}
L_{\mathrm{bin}, (1)}&=\int_{0}^{\infty} \frac{{e^{-\lambda}}{\lambda}^{k} }{k!} \cdot \left[  \mathrm{GPG}_1(\lambda_{1})  \ast \ldots \ast  \mathrm{GPG}_N(\lambda_{N})\right](\lambda) \ d\lambda \\
&=\sum\limits_{k_{mc,1}=0}^{\infty} \ldots \sum\limits_{k_{mc,M}=0}^{\infty} \mathrm{PG}_1 \cdot \mathrm{PG}_M \cdot \int_{0}^{\infty} \frac{{e^{-\lambda}}{\lambda}^{k} }{k!} \cdot \left[  \mathrm{G}_1(\lambda_{1})  \ast \ldots \ast  \mathrm{G}_N(\lambda_{N})\right](\lambda) \ d\lambda \\
&=\sum\limits_{k_{mc,1}=0}^{\infty} \ldots \sum\limits_{k_{mc,M}=0}^{\infty} \mathrm{PG}_1 \cdot \mathrm{PG}_M \cdot \sum_{\substack{k_1 + \ldots + k_M = k\\k_j\geq 0}}  \prod\limits_{j=1}^{M} P_{\mathrm{PG}}(k_j;k_{mc,j}+\alpha_j^{*},1/w_j)  \\
&=\sum_{\substack{k_1 + \ldots + k_M = k\\k_j\geq 0}}  \prod\limits_{j=1}^{M} \sum\limits_{k_{mc,j}=0}^{\infty}P_{\mathrm{PG}}(k_j;k_{mc,j}+\alpha_j^{*},1/w_j) \cdot P_{\mathrm{PG,i}(k_{mc,j};\langle k_{mc} \rangle_j + \beta_j^*, 1/v_j)} \label{eq:pgpg_combinatorial}
\end{align}
where we first pull out all the infinite sums, then rewrite the integral over the convolution of gamma factors with a Poisson factor in combinatorial form (see appendix \ref{appendix:general_formulas}) and then again combine each summation with the respective terms. The result is a combinatorial sum over $\mathrm{PGPG}$ distributions, each of which we calculated the result in the previous section. We can continue and replace every $\mathrm{PGPG}$ distribution in eq. \ref{eq:pgpg_combinatorial} by its respective contour integral (eq. \ref{eq:pgpg_contour}), which results in

\begin{align}
L_{\mathrm{bin}, (1)} &= \sum_{\substack{k_1 + \ldots + k_M = k\\k_j\geq 0}}  \prod\limits_{j=1}^{M} \sum\limits_{k_{mc,j}=0}^{\infty}P_{\mathrm{PG}}(k_j;k_{mc,j}+\alpha_j^{*},1/w_j) \cdot P_{\mathrm{PG,i}(k_{mc,j};\langle k_{mc} \rangle_j + \beta_j^*, 1/v_j)} \label{eq:exp_value_over_combinatorial} 
\\
&= \sum_{\substack{k_1 + \ldots + k_M = k\\k_j\geq 0}}  \prod\limits_{j=1}^{M} 
K_j \cdot \frac{1}{2\pi i}\oint\limits_{|z_j|=\epsilon} \frac{(1+z_j)^{k_j+\alpha_j^*-1} }{ z_j^{k_j+1}} \cdot \Big(\frac{1}{1-C_j(1+z_j)}\Big)^{\langle k_{mc}\rangle_j+\beta_j^*} dz_j \label{eq:before_egorychev}
\end{align}
where we use $\alpha_j^*=\frac{\langle k_{mc,j} \rangle \cdot \alpha}{N}$ and $\beta_j^*=\frac{\langle k_{mc,j} \rangle \cdot \beta}{N}$ to share the nuisance parameter among groups of events with different weights and the definition of $K_j$ and $C_j$ is the same as $K$ and $C$ for given a weight group in the previous section. Again we use $\langle k_{mc} \rangle$ to denote observed number of MC events, and $k_{mc}$ as a variable to sum over. We can now solve the constrained sum over $k_j$ using Egorychev's rules for constrained summation indices \cite{Egorychev1984}. The constraint here is $\sum k_j = k$, which is encoded with complex weight variables $\tau$ of equal magnitude. Equation eq. \ref{eq:before_egorychev} with these weight variables $\tau$ and an extra outer contour integral then can be re-written as
\begin{align}
L_{\mathrm{bin}, (1)}&=\oint\limits_{|\tau|=\epsilon}^{} \frac{1}{\tau^{k+1}} \stackrel[k_1=0 \ldots k_M=0]{\infty \ldots \infty}{\sum}  \prod\limits_{j=1}^{M} 
K_j \cdot \frac{1}{2\pi i}\oint\limits_{|z_j|=\epsilon} \frac{(1+z_j\cdot\tau)^{k_j+\alpha_j^*-1} }{ z_j^{k_j+1}} \cdot \Big(\frac{1}{1-C_j(1+z_j\cdot\tau)}\Big)^{\langle k_{mc}\rangle_j+\beta_j^*} dz_j \\
&=
\begin{aligned}
\oint\limits_{|\tau|=\epsilon}^{} &\frac{1}{\tau^{k+1}} \stackrel[k_1=0 \ldots k_M=0]{\infty \ldots \infty}{\sum}  \prod\limits_{j=1}^{M} 
K_j^{'}\cdot  \left(\frac{1}{1+1/w_j}\right)^{k_j} \\ \cdot &\frac{1}{2\pi i}\oint\limits_{|z_j|=\epsilon}
\underbrace{\Big(\frac{1}{1-C_j(1+z_j\cdot\tau)}\Big)^{\langle k_{mc}\rangle_j+\beta_j^*} \cdot (1+z_j\cdot\tau)^{\alpha_j^*-1}}_{\varphi} \cdot (\underbrace{1+z_j\cdot\tau}_{f})^{k_j} \cdot  z_j^{-k_j-1} dz_j d \tau 
\end{aligned} \\ &=\oint\limits_{|\tau|=\epsilon}^{} \frac{1}{\tau^{k+1}}   \prod\limits_{j=1}^{M} 
K_j^{'} \cdot \frac{(1-E_j \cdot \tau)^{\langle k_{mc}\rangle_j+\beta_j^*-\alpha_j^*}}{(1-C_j - E_j\cdot \tau)^{\langle k_{mc}\rangle_j+\beta_j^*}} d \tau \\
&=\left(\prod\limits_{j=1}^{M}  K_j^{'} \cdot (-E_j)^{-\alpha_j^*}\right) \oint\limits_{|\tau|=\epsilon}^{} \frac{1}{\tau^{k+1}}   \prod\limits_{j=1}^{M} 
\cdot \frac{(\tau-1/E_j)^{\langle k_{mc}\rangle_j+\beta_j^*-\alpha_j^*}}{(\tau - (1-C_j)/E_j )^{\langle k_{mc}\rangle_j+\beta_j}} d \tau \label{eq:contour_int_gen1} \\
&= \left(\prod\limits_{j=1}^{M}  K_j^{'} \cdot \left(\frac{1}{1-C_j}\right)^{\langle k_{mc}\rangle_j+\beta_j}\right) \cdot \Delta_k \\
&\stackrel[\substack{\alpha=0 \\ \beta=0 \\ v_j=1}]{}{=} \left(\prod\limits_{j=1}^{M} \cdot \left(\frac{1+w_j}{1+2w_j}\right)^{\langle k_{mc}\rangle_j}\right) \cdot \Delta_k \label{eq:final_generalized_formula}
\end{align} 
using $K_j \equiv K_j^{'}\cdot E_j^{k_j}$, $E_j=\frac{1}{1+1/w_j}$, $C_j=\frac{v_j}{(1+v_j)(1+w_j)}$ and $K_j^{'}=(\frac{1}{1+w_j})^{\alpha^*} \cdot (\frac{1}{1+v_j})^{\langle k_{mc} \rangle+\beta_j^*}$ from the definitions in the previous section. The terms we call $f$ and $\varphi$ are highlighted to indicate the structure used for the main theorem described in \cite{Egorychev1984} (p. 19, 1.4.2). The term $\Delta_k$ represents an efficient solution \cite{Ma2014} of the contour integral in eq. \ref{eq:contour_int_gen1}, and is written as
\begin{align}
\Delta_k&=\frac{1}{k}\sum\limits_{i=1}^{k} \left[\left(\sum\limits_{j=1}^{M} (\langle k_{mc,j} \rangle + \beta_j^*) \cdot \left(\frac{E_j}{1-C_j}\right)^i - \Big(\langle k_{mc,j} \rangle + \beta_j^* -\alpha_j^*\Big) \cdot {E_j}^i \right) \Delta_{k-i}\right] \\
&=\frac{1}{k}\sum\limits_{i=1}^{k} \left[\left(\sum\limits_{j=1}^{M} \left(\langle k_{mc,j} \rangle + \beta_j^* \right)\cdot \left(\frac{w_j (1+v_j)}{1+w_j(1+v_j)}\right)^i - \Big(\langle k_{mc,j} \rangle + \beta_j^* -\alpha_j^* \Big) \cdot \left({\frac{1}{1+1/w_j}}\right)^i \right) \Delta_{k-i}\right] \\
&\stackrel[\substack{\alpha=0 \\ \beta=0 \\ v_j=1}]{}{=}\frac{1}{k}\sum\limits_{i=1}^{k} \left[\left(\sum\limits_{j=1}^{M} \langle k_{mc,j} \rangle \cdot \left(\frac{2}{2+1/w_j}\right)^i - \langle k_{mc,j} \rangle  \cdot \left({\frac{1}{1+1/w_j}}\right)^i \right) \Delta_{k-i}\right] 
\end{align}
with $\Delta_0=1$. One can also write the sum going over all individual events $N$ instead over $M$ groups of events that share a similar weight, which results in $k_{mc,j}=1$ for each summand given by the last equal sign. The nuisance parameters $\alpha_j^*=\frac{\langle k_{mc,j} \rangle \cdot \alpha}{N}$ and $\beta_j^*=\frac{\langle k_{mc,j} \rangle \cdot \beta}{N}$ are typically set to zero since $\alpha=\beta=0$ is a natural prior choice (see also \cite{Gluesenkamp2018}). The parameter $v_j$ is further set to unity, because it came about as the scaling parameter of the gamma distribution in the Poisson gamma mixture describing pure counts. The general case of convolutions of arbitrarily mixed gamma distributions and gamma-Poisson-gamma mixtures is described in appendix \ref{appendix:general_formulas}. 
\section{Derivation generalization (3)}
\label{appendix:gen_3}

\subsection{Single Source Dataset}

For a single source dataset we have a single gamma factor that approximates the CPD similar to the description in section \ref{sec:say_llh}. The corresponding integral looks like
\begin{align}
\mathrm{PGG}(k)&=\int\limits_{0}^{\infty} \int\limits_{0}^{\infty} \frac{e^{-k} \cdot \lambda^k}{k!} G(\lambda;\mu\cdot Q, \beta) \cdot \mathrm{G}(\mu;k_{mc},\gamma) \ d \lambda \ d\mu = \int\limits_{0}^{\infty} \mathrm{PG}(k;\mu\cdot Q, \beta) \cdot \mathrm{G}(\mu;k_{mc},\gamma)  \ d\mu \label{eq:single_gammagamma_first} \\
&=\int\limits_{0}^{\infty}
\frac{\Gamma(\mu\cdot Q+k)}{\Gamma(\mu\cdot Q) k!} \cdot {\underbrace{\left(\frac{\beta}{1+\beta}\right)}_{\equiv A}}^{\mu \cdot Q} \cdot {\underbrace{\left(\frac{1}{1+\beta}\right)}_{\equiv B}}^k \cdot \mathrm{G}(\mu;k_{mc},\gamma) \ d\mu \label{eq:single_gammagamma_second}
\end{align}
where the first gamma factor in eq. \ref{eq:single_gammagamma_first} represents the CPD approximation of the source dataset with $Q=\frac{1}{k_{mc}}\frac{(\sum_j w_j)^2}{\sum_j w_j^2}$, $\beta=\frac{\sum_j w_j}{\sum_j w_j^2}$, and the second gamma factor marginalizes over the unknown rate where we leave a scaling parameter $\gamma$ for generality. We will later set $\gamma=1$. Additionally we define $A=\frac{\beta}{1+\beta}$ and $B=\frac{1}{1+\beta}$.
 We can also use the fact that $\frac{\Gamma(x+k)}{\Gamma(x)}=\sum\limits_{n=0}^{k} {k \brack n} x^n$ \cite{Graham1989} where $ {k \brack n}$ are the unsigned Stirling numbers of the first kind and reformulate eq. \ref{eq:single_gammagamma_second} as
\begin{align}
\mathrm{PGG}(k)&=\int\limits_{0}^{\infty}
\frac{\Gamma(\mu\cdot Q+k)}{\Gamma(\mu\cdot Q) k!} \cdot {\left(\frac{\beta}{1+\beta}\right)}^{\mu \cdot Q} \cdot {\left(\frac{1}{1+\beta}\right)}^k \cdot \mathrm{G}(\mu;k_{mc},\gamma) \ d\mu \\
&=\int\limits_{0}^{\infty}
\frac{1}{ k!} \sum\limits_{n=0}^{k} {k \brack n} (\mu Q)^n  \cdot {A}^{\mu \cdot Q} \cdot B^k \cdot \frac{e^{-\gamma \cdot \mu} \cdot \mu^{k_{mc}-1}}{\Gamma(k_{mc})} \cdot \gamma^{k_{mc}} \ d\mu \\
&= B^k \cdot \gamma^{k_{mc}}\sum\limits_{n=0}^{k} 
\frac{1}{ k!} \cdot Q^n  {k \brack n} 
\frac{\Gamma(k_{mc}+n)}{\Gamma(k_{mc})(\gamma-Q\mathrm{ln}(A))^{k_{mc}+n}}
\int\limits_{0}^{\infty}  \frac{e^{-\mu(\gamma-Q\mathrm{ln}(A))} \cdot \mu^{k_{mc}+n-1} }{\Gamma(k_{mc}+n)} (\gamma-Q\mathrm{ln}(A))^{k_{mc}+n}  \ d\mu \\
&=B^k \cdot \gamma^{k_{mc}}\sum\limits_{n=0}^{k} 
\frac{1}{ k!} \cdot Q^n  {k \brack n} 
\frac{\Gamma(k_{mc}+n)}{\Gamma(k_{mc})(\gamma-Q\mathrm{ln}(A))^{k_{mc}+n}}
\end{align}

\section{Overview of some generic formulas}
\label{appendix:general_formulas}
The solutions of generalization (1) \ref{sec:generalization_a} can be written in various forms which we summarize in the following. We abbreviate the Poisson distribution as $\mathrm{P}$, the gamma distribution as $\mathrm{G}$, the gamma-Poisson mixture as $\mathrm{GP}$, and the Negative-Binomial distribution as $\mathrm{NB}$. When we marginalize the shape parameter of a gamma distribution with a $\mathrm{GP}$ distribution, we abbreviate that as $\mathrm{GPG}$, and a marginalization of a $\mathrm{GP}$ distribution shape parameter with another $\mathrm{GP}$ distribution is abbreviated as $\mathrm{GPGP}$.

\subsection{Poisson-gamma mixture form}

\subsubsection{Simple generalization}
The PMF of the generalized Poisson-gamma mixture has already derived in \cite{Gluesenkamp2018}. In \cite{Vellaisamy2009} it was shown that the convolution of Poisson-gamma mixtures is actually describing the same distribution, and so we have

\begin{align}
P_{gen}(k; \bm{\alpha},\bm{\beta}) &= [\mathrm{PG}(k_1;\alpha_1,\beta_1) \ast \ldots \ast \mathrm{PG}(k_N;\alpha_N,\beta_N)](k) \label{eq:pg_sums} \\ 
&= \int\limits_{0}^{\infty} \mathrm{P}(k;\lambda) \cdot \left[\mathrm{G}(\lambda_1;\alpha_1, \beta_1) \ast \ldots \ast \mathrm{G}(\lambda_N;\alpha_N, \beta_N)\right](\lambda) \ d\lambda  \\
&= \stackrel[\sum_i k_i=k, \ k_i \geq 0]{}{\sum}  \prod_i \frac{ \Gamma(k_i+\alpha_i)  }{k_i!\cdot \Gamma(\alpha_i)} \cdot \beta_i^{\alpha_i} \cdot \left(\frac{1}{1+\beta_i} \right)^{k_i+\alpha_i} \label{eq:poisson_gamma_mixture_simple_combinatorial} \\
&= \stackrel[\sum_i k_i=k, \ k_i \geq 0]{}{\sum}  \prod_i \mathrm{PG}(k_i;\alpha_i,\beta_i)  \\
&=  D_k(\bm{\alpha},\bm{\beta}) \cdot \prod_i \left(\frac{\beta_i}{1+\beta_i}\right)^{\alpha_i}
\end{align}
with iterative definition
\begin{align} D_k(\bm{\alpha},\bm{\beta})=\frac{1}{k}\sum\limits_{j=1}^{k} \left[\left(\sum\limits_{i=1}^{N} \alpha_i \cdot {\frac{1}{1+\beta_i}}^j \right) D_{k-j}\right] \ \ \mathrm{and} \ D_0=1. 
\end{align}
The combinatorial form (eq. \ref{eq:poisson_gamma_mixture_simple_combinatorial}) has been derived in \cite{Gluesenkamp2018} and in \cite{Vellaisamy2009} in the different context of a convolution of Poisson-gamma mixtures. However, the calculation based on the iterative sum (eq. \ref{eq:poisson_gamma_mixture_simple_iterative}) is much more efficient. The convolution of Poisson-gamma mixtures indicated in eq. \ref{eq:pg_sums} is an important step in the efficient calculation derived in appendix \ref{appendix:gen_3}. This distribution is very rich in structure, and its relationship to the Carlson-R function \cite{Carlson1963} and Lauricella function $F_D$ \cite{Lauricella1893} have been explored in \cite{Gluesenkamp2018}. The iterative formula has substantial computational advantages in various applications of these special functions.

\subsubsection{Extended generalization}

In appendix \ref{appendix:gen_1} we derived a generalization of the extended Poisson gamma mixture by taking a further expectation value (eq. \ref{eq:exp_value_over_combinatorial}) with respect another Poisson-gamma mixture. Here, we proceed more generally and only take the expectation value of $M$ Poisson-gamma terms while $N$ standard terms remain. Starting with the combinatorial expression similar to eq. \ref{eq:exp_value_over_combinatorial} we can write
\begin{align}
P_{gen}(k; \bm{\alpha},\bm{\beta}, \bm{\gamma}, \bm{\delta}, \bm{\varepsilon}) 
&= \stackrel[\substack{\sum_i k_i+\sum_j k_j=k \\  k_i \geq 0 ,  k_j \geq 0}]{}{\sum}  \prod\limits_{i=1}^{N} \mathrm{PG}(k_i;\alpha_i,\beta_i) \prod\limits_{j=1}^{M}  \sum\limits_{t=0}^{\infty} \mathrm{PG}(k_j;t,\gamma_j) \mathrm{PG}(t;\delta_j,\varepsilon_j)  \label{eq:extended_pg_gamma_generalization_1}  \\ 
&= \stackrel[\substack{\sum_i k_i+\sum_j k_j=k \\  k_i \geq 0 ,  k_j \geq 0}]{}{\sum}  \prod\limits_{i=1}^{N} \mathrm{PG}(k_i;\alpha_i,\beta_i) \prod\limits_{j=1}^{M}  \mathrm{PGPG}(k_j;\gamma_j, \delta_j, \varepsilon_j)\\
&= 
\begin{aligned}
\int\limits_{0}^{\infty} \mathrm{P}(k;\lambda) \cdot \bigg[\mathrm{G}(\lambda_1;\alpha_1, \beta_1) \ast \ldots \ast \mathrm{G}(\lambda_N;\alpha_N, \beta_N) \\ \ast \mathrm{GPG}(\lambda_{1^*};\gamma_1, \delta_1, \varepsilon_1) \ast \ldots \ast &\mathrm{GPG}(\lambda_{M^*};\gamma_M, \delta_M, \varepsilon_M) \bigg](\lambda) \ d\lambda 
\end{aligned}
\\
&=
\begin{aligned}
\bigg[\mathrm{PG}(k_1;\alpha_1,\beta_1) \ast \ldots \ast \mathrm{PG}(k_N;\alpha_N,\beta_N) \\ \ast \mathrm{PGPG}(k_{1^*};\gamma_1,\delta_1,\varepsilon_1) \ast \ldots \ast  &\mathrm{PGPG}(k_{M^*};\gamma_M,\delta_M,\varepsilon_M)\bigg](k) 
\end{aligned}  \\ 
&=\Delta_k(\bm{\alpha},\bm{\beta}, \bm{\gamma}, \bm{\delta}, \bm{\varepsilon}) \cdot \prod\limits_{i=1}^{N} \left(\frac{\beta_i}{1+\beta_i}\right)^{\alpha_i} \prod\limits_{j=1}^{M} \left(\frac{\varepsilon_j \cdot (\gamma_j+1)}{\varepsilon_j \cdot (\gamma_j+1)+1}\right)^{\delta_j}
\end{align}
\begin{align}
\Delta_k(\bm{\alpha},\bm{\beta}, \bm{\gamma}, \bm{\delta}, \bm{\varepsilon})=\frac{1}{k}\sum\limits_{j=1}^{k} \left[\left(\sum\limits_{i=1}^{N} \alpha_i \cdot \left({\frac{1}{1+\beta_i}}\right)^j + \sum\limits_{h=1}^{M} \delta_h \cdot \bigg[\left(\frac{1+\varepsilon_h}{1+\varepsilon_h \cdot(1+\gamma_h)}\right)^j-\left({\frac{1}{1+\gamma_h}}\right)^j\bigg]  \right) \Delta_{k-j}\right] \ \ \mathrm{and} \ \Delta_0=1
\end{align}
where we have have introduced additional M-dimensional vectors $\bm{\gamma}$, $\bm{\delta}$, $\bm{\varepsilon}$ that parametrize the additional $M$ marginalized factors resulting in the $\mathrm{PGPG}$ or $\mathrm{GPG}$ distributions, depending how we write it. If we put $M=N$, $\delta_i=\langle k_{mc,i} \rangle$, $\gamma_i=\beta_i=1/w_i$ and $\varepsilon_i=1$ we obtain eq. \ref{eq:final_generalized_formula} as a special case.

\subsection{Negative Binomial form}

Poisson-gamma mixtures are often re-parametrized as negative binomial distributions, which can be obtained by setting $\beta=\frac{1-p}{p}$ and using $\alpha=r$. For completion we give the simple and extended negative-binomial generalizations below. 

\subsubsection{Simple generalization}

\begin{align}
\mathrm{NB}_{\mathrm{simple}}(k; \bm{r},\bm{p}) &=  \int\limits_{0}^{\infty} \mathrm{P}(k;\lambda) \cdot \left[\mathrm{G}(\lambda_1;r_1, \frac{1-p_1}{p_1}) \ast \ldots \ast \mathrm{G}(\lambda_N;r_N, \frac{1-p_N}{p_N})\right](\lambda) \ d\lambda \\
&= D_k(\bm{r},\bm{p}) \cdot \prod_i \left(1-p_i\right)^{r_i} \label{eq:nb_simple_iterative}
\end{align}
with iterative definition
\begin{align} D_k(\bm{r},\bm{p})=\frac{1}{k}\sum\limits_{j=1}^{k} \left[\left(\sum\limits_{i=1}^{N} r_i \cdot {p_i}^j \right) D_{k-j}\right] \ \ \mathrm{and} \ D_0=1. 
\end{align}
If all $r_i$ expect one are zero, which effectively means there is only one pair of $r_i$ and $p_i$, we obtain the standard negative binomial distribution as a special case.
\subsubsection{Extended generalization}

In order to obtain an extended negative binomial distribution that is sensible, we take eq. \ref{eq:extended_pg_gamma_generalization_1} and set $\beta_i=\frac{1-p_i}{p_i}$, $\gamma_i=\frac{1-p^*_i}{p^*_i}$, $\alpha_i=r_i$, $\delta_i=r^*_i$ and $\varepsilon_i=1$. The result gives more variance to the gamma factors with $p^*_i$ and $r^*_i$ which are marginalized with an extra Poisson-gamma factor. Using unity for $\varepsilon_i$ is the natural choice that is also used for the application in eq. \ref{eq:final_generalized_formula}. We obtain
\begin{align}
\mathrm{NB}_{\mathrm{ext}}(k; \bm{r},\bm{p},\bm{r^*},\bm{p^*}) &=  
\begin{aligned}
\int\limits_{0}^{\infty} \mathrm{P}(k;\lambda) \cdot \bigg[\mathrm{G}(\lambda_1;r_1, \frac{1-p_1}{p_1}) \ast \ldots \ast \mathrm{G}(\lambda_N;r_N, \frac{1-p_N}{p_N}) \\ \ast \mathrm{GPG}(\lambda_{1^*};\frac{1-p^*_1}{p^*_1}, r^*_1, 1) \ast \ldots \ast &\mathrm{GPG}(\lambda_{M^*};\frac{1-p^*_M}{p^*_M}, r^*_M, 1) \bigg](\lambda) \ d\lambda 
\end{aligned}  \\ 
&= \Delta_k(\bm{r},\bm{p}, \bm{r^*}, \bm{p^*}) \cdot \prod\limits_{i=1}^{N} \left(1-p_i\right)^{r_i} \prod\limits_{j=1}^{M} \left(\frac{1}{ 1+p^*_j}\right)^{r^*_j} \label{eq:nb_extended_iterative}
\end{align}
with iterative definition
\begin{align} \Delta_k(\bm{r},\bm{p}, \bm{r^*}, \bm{p^*})=\frac{1}{k}\sum\limits_{j=1}^{k} \left[\left(\sum\limits_{i=1}^{N} r_i \cdot p_i^j + \sum\limits_{h=1}^{M} r^*_h \cdot \bigg[\left(\frac{p^*_h}{1+p^*_h}\right)^j-(p^*_h)^j\bigg]  \right) \Delta_{k-j}\right] \ \ \mathrm{and} \ \Delta_0=1.
\end{align}

\section{Supplementary figures}
\label{appendix:suppl_plots}

\begin{figure}
	\centering
	{\includegraphics{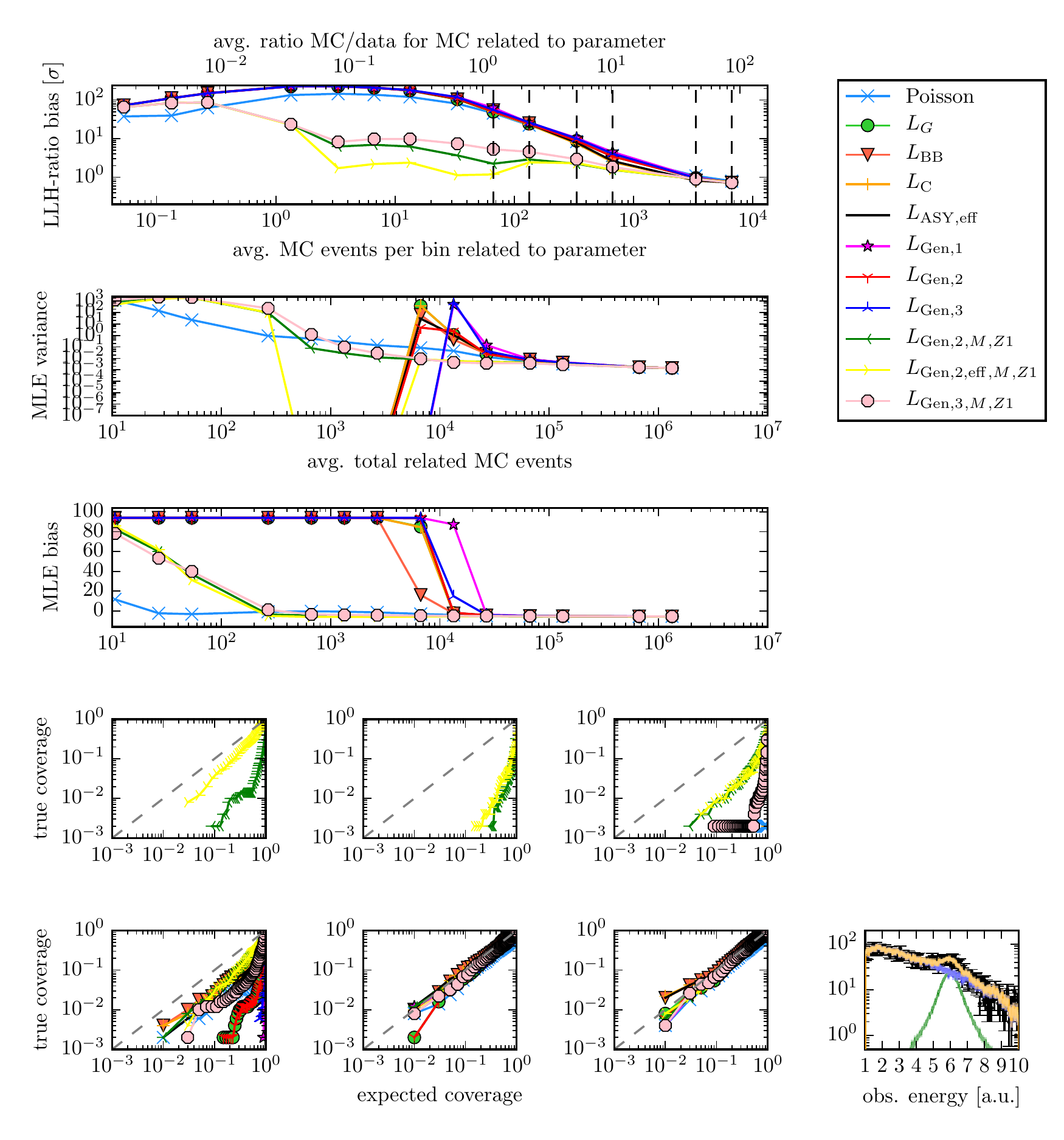}}
	\caption{Bias of the log-likelihood ratio $\lambda$ and coverage in dependence of the signal dataset statistics. The background statistics stays the same. The parameter of interest is the normalization of the peak in the signal dataset. $Z1$ denotes strategy 1 for "zero MC", $Z2$ strategy 2 for "zero MC", $U$ denotes more "unbiased" weight variance estimation and $M$ denotes shifted mean based on average MC events in all bins. The coverage is shown at 6 statistic levels (in order) which are indicated as dashed vertical lines in the upper plot. In the lower right plot a particular data realization and a MC realization at the highest statistic level is shown for the observable space. The non-standard likelihoods from other publications include $L_{\mathrm{ASY,eff}}$\cite{Argueelles2019},  $L_{\mathrm{BB}}$ \cite{Barlow1993}, $L_{\mathrm{C}}$ \cite{Chirkin2013b} and  $L_{\mathrm{G}}$\cite{Gluesenkamp2018}.} \label{fig:mid_2_bias}
\end{figure}

\begin{figure}
	\centering
	{\includegraphics{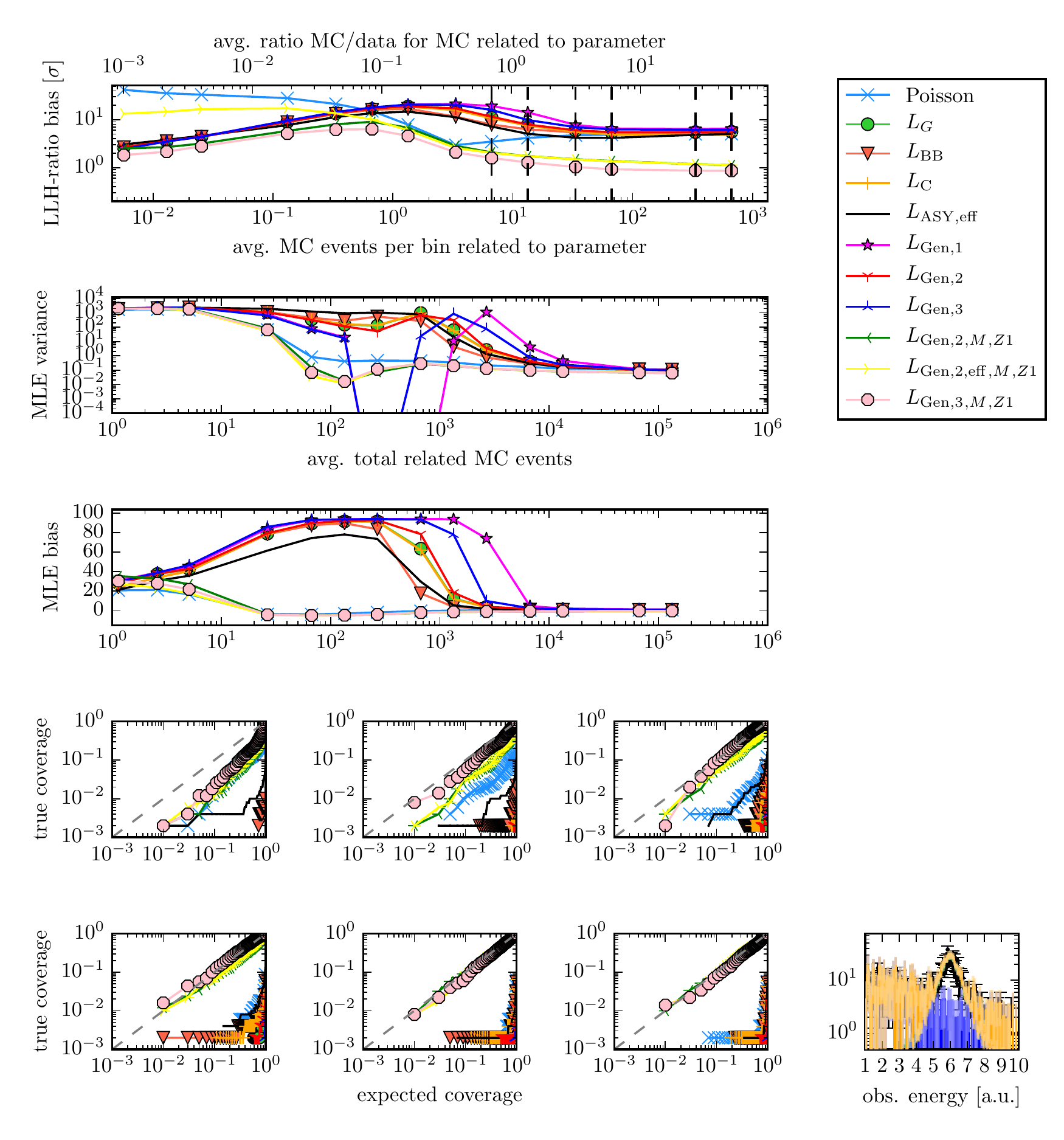}}
	\caption{Bias of the log-likelihood ratio $\lambda$ and coverage in dependence of the signal dataset statistics. Signal and background statistics increase together. The parameter of interest is the normalization of the peak in the signal dataset. $Z1$ denotes strategy 1 for "zero MC", $Z2$ strategy 2 for "zero MC", $U$ denotes more "unbiased" weight variance estimation and $M$ denotes shifted mean based on average MC events in all bins. The coverage is shown at 6 statistic levels (in order) which are indicated as dashed vertical lines in the upper plot. In the lower right plot a particular data realization and a MC realization at the highest statistic level is shown for the observable space. The non-standard likelihoods from other publications include $L_{\mathrm{ASY,eff}}$\cite{Argueelles2019},  $L_{\mathrm{BB}}$ \cite{Barlow1993}, $L_{\mathrm{C}}$ \cite{Chirkin2013b} and  $L_{\mathrm{G}}$\cite{Gluesenkamp2018}.} \label{fig:highbg_lowpeak_bias}
\end{figure}

\begin{figure}
	\centering
	{\includegraphics{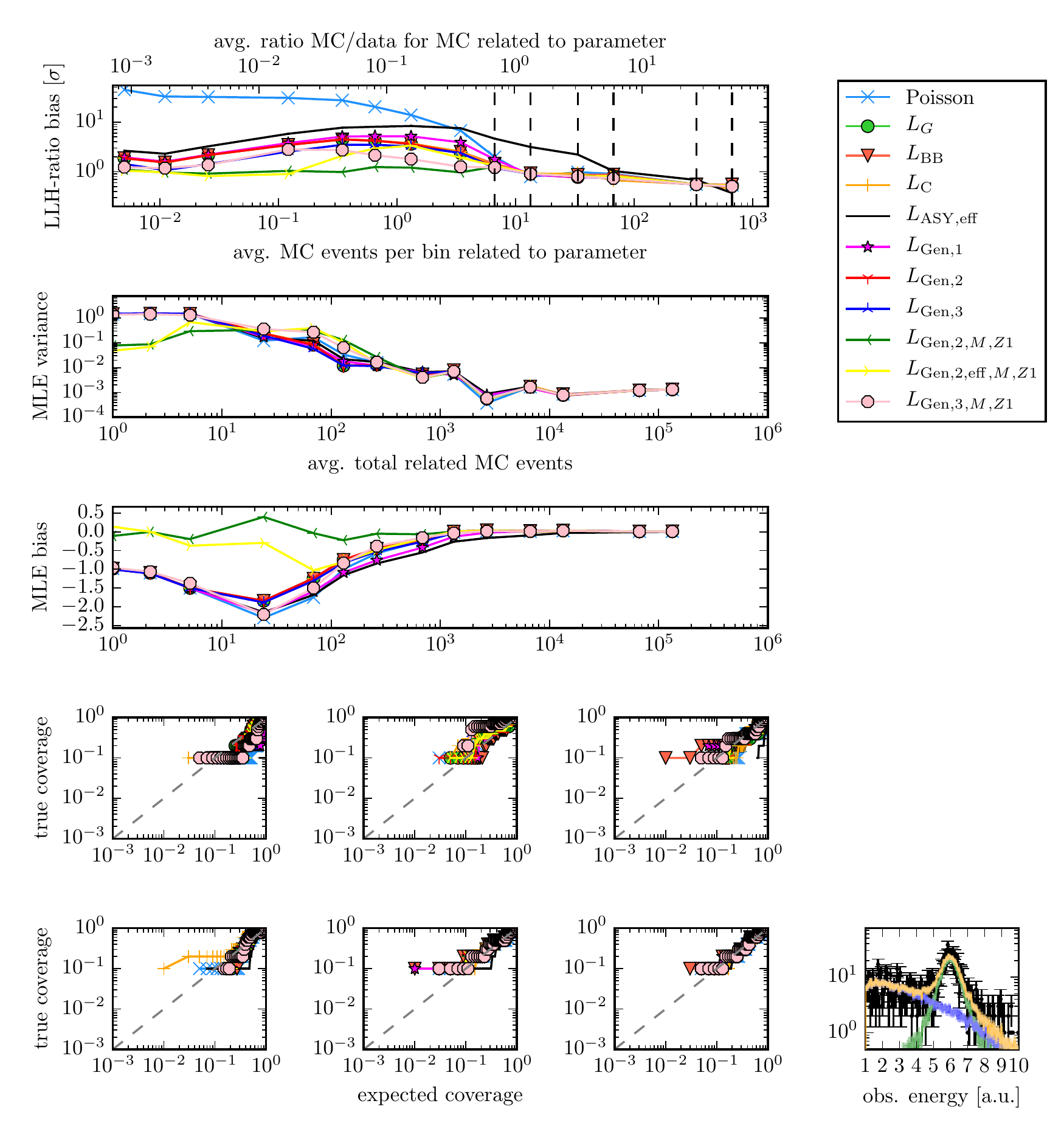}}
	\caption{Bias of the log-likelihood ratio $\lambda$ and coverage in dependence of the background dataset statistics. The signal and background statistics increase together. The parameter of interest is the spectral index of the background dataset. $Z1$ denotes strategy 1 for "zero MC", $Z2$ strategy 2 for "zero MC", $U$ denotes more "unbiased" weight variance estimation and $M$ denotes shifted mean based on average MC events in all bins. The coverage is shown at 6 statistic levels (in order) which are indicated as dashed vertical lines in the upper plot. In the lower right plot a particular data realization and a MC realization at the highest statistic level is shown for the observable space. The non-standard likelihoods from other publications include $L_{\mathrm{ASY,eff}}$\cite{Argueelles2019},  $L_{\mathrm{BB}}$ \cite{Barlow1993}, $L_{\mathrm{C}}$ \cite{Chirkin2013b} and  $L_{\mathrm{G}}$\cite{Gluesenkamp2018}.} \label{fig:sifit_bias}
\end{figure}

\bibliographystyle{unsrt}
\bibliography{generalized_pg_mixtures}

\end{document}